\documentclass[prc,aps,tighten,preprint,showpacs]{revtex4}
\pdfoutput=1
\usepackage{graphicx}
\usepackage{dcolumn}
\usepackage{bm}
\usepackage{slashed}
\usepackage{float}
\usepackage{amsmath,amssymb}
\usepackage{hyperref}
\usepackage{color}
\usepackage{grffile}
\usepackage{fancybox}
\usepackage[usenames,dvipsnames]{xcolor}
\newcommand{\sh}[1]{#1\hskip -6pt  / }
\newcommand{\shl}[1]{#1\hskip -4pt  / }

\graphicspath{ {figures/} }

\begin{document}

\title{Electron scattering from a deeply bound nucleon on the light-front}

\author{Frank Vera and Misak M. Sargsian}
\affiliation{Department of Physics, Florida International University, Miami,
Florida 33199, USA}
  
\date{\today}

\begin{abstract}
We calculate the cross section of the electron scattering from a bound nucleon within light-front approximation.
The advantage of this approximation is the possibility of 
systematic account for the off-shell effects which become 
essential in high energy electro-nuclear processes aimed at  probing  the nuclear structure at small distances.
We derive a new dynamical parameter which allows to control the extent of the "off-shellness" of electron -  bound-nucleon 
electromagnetic current for different regions of momentum transfer and initial light-cone momenta of the bound nucleon.
The derived cross section is compared with the results of other approaches in treating the off-shell effects in 
electron-nucleon scattering.
\end{abstract}
\maketitle

\section{Introduction}

High energy electro-nuclear  processes  ranging from inclusive $A(e,e^\prime)X$ to  the double 
$A(e,e^\prime N_f)X$ and triple $A(e,e^\prime N_f,N_r)X$   coincidence reactions, in which  $e^\prime$ is the scattered electron, 
$N_f$ and $N_r$ are struck and recoil nucleons are the main processes used to probe the short-range structure of nuclei.
During the last two decades a  multitude of dedicated  experiments have been performed that significantly 
advanced  our understanding of the dynamics of short-range nuclear correlations (for recent reviews on this subject see 
Refs.\cite{arnps,Hen:2016kwk,dreview,srcprog,Atti:2015eda,srcrev}).  All of these experiments were performed at  quasi-elastic kinematics in which 
electrons are scattered off the  deeply bound nucleon producing a struck nucleon ($N_f$) in the final state. 
The observed  experimental signatures were in agreement with the expectations that  deeply bound nucleons emerge from 
short-range nucleon-nucleon correlations~(SRCs). 
These signatures included the onset of  scaling for  the inclusive  $A(e,e^\prime)X$  cross section ratios of nucleus, A  to 
the deuteron or $^3He$\cite{FSDS93,Kim1,Kim2,Fomin:2011ng}, strong angular correlation between 
momenta of struck, $N_f$ and recoil, $N_r$ nucleons\cite{EIP2,EIP3}  as well as significant dominance 
of the $pn$ correlations\cite{isosrc,EIP4,newprops,Hen:2014nza}  in the domain of 2N SRCs.  

The next stage of  SRC studies requires the exploration of quantitative properties of the nuclear structure 
describing nucleons in the SRC.   This research can be both experimental -  performing  extraction of nuclear spectral and decay functions 
in the region of high momentum and removal energy of the struck nucleon or theoretical-  by modeling
these structure functions~(see e.g. \cite{Artiles_Sargsian-multisrc1}) and predicting electroproduction cross sections in
large missing  momentum and removal energy kinematics.   One of the outstanding problems in such 
research is the  understanding of the reaction mechanism and 
final state interaction~(FSI) effects associated with the electron scattering from a  deeply bound nucleon in 
the nucleus. 

During the last two decades significant efforts have been made  in the calculation of FSI effects in high $Q^2$  electro-nuclear 
processes (see e.g. Refs.\cite{gea,ms01,CiofidegliAtti:2004jg,Jeschonnek:2008zg,Laget04,eheppn1,eheppn2,Sargsian:2009hf,edex,disfsirev}).  
One of the approaches,  referred to as generalized eikonal approximation\cite{gea,Sargsian:2009hf},    
selfconsiestently treated the relativistic effects associated with the large momentum  of bound nucleon 
involved in the reaction, as such these approach provided a  theoretical framework for 
calculating FSI effects relevant to studies of the nuclear structure at short distances.

However not much theoretical attention is given currently  to the studies of the reaction mechanism of  elastic scattering 
from the  high momentum bound nucleon in the nucleus.    The problem of the proper description of  electromagnetic scattering from 
deeply bound nucleon in the nucleus was realized in 1980's with the  advent of the  intermediate energy $A(e,e^\prime N_f)$ 
experiments at SACLAY\cite{SACLAY1,SACLAY2} and  NIKHEF\cite{NIKHEF}.  
The first approaches  in describing  electron-deeply bound nucleon scattering  were based on different methods of 
interpreting the spinor of the bound (off-shell) nucleon.  In one of the earlier models\cite{Mougey} the on-shell nucleon spinors
were used with the mass estimated as $m_N^{* \ 2} = E^2 - p^2$.  
Currently the most popular model is that of de Forest\cite{deForest(1983)} in which 
different expressions for the $eN_{bound}$ cross sections are  obtained based on the different assumptions for 
effective $\gamma^*N_{bound}$ vertices with on-shell spinors used for the bound nucleon.  
No preference is given to any of the considered eight expressions of  the $eN_{bound}$ cross section and as such 
these approximations allowed to check uncertainty due to the binding effects  
rather than calculating their actual  values.  Such an approach was characteristic to the intermediate energy (few hundred MeV of 
incoming beam energy) scattering processes in which no  small parameters existed in treating the strong binding effects  in 
nucleon electromagnetic current.

The situation has recently changed
 with the emergence of high energy and momentum transfer  $eA$ experiments 
(see e.g. \cite{arnps,dreview,srcprog}) in which deeply bound nucleons in the nucleus are probed with high $Q^2$ virtual photons 
producing
final nucleons with momenta above $few$~GeV/c region. 
The high energy nature of the scattering process allows for
important simplifications in describing the scattering process  similar to those in hadronic physics.
One of the main characteristics of high energy 
scattering  is that the process evolves along the light-front (see e.g. \cite{Feynman,KS(1970),LB(1980),FS81,BPP(1997)}) which makes the 
light-cone the most natural reference frame to describe the reaction. The important advantage of such description is the suppression of 
the negative energy contribution in the propagator of bound nucleon as well as
 the  possibility of identifying the "good" component of  electromagnetic current  for which the off-shell effects are minimal.  There have been 
several  extensive studies of nuclear dynamics on light-front (see e.g.  Refs.\cite{FS81,FS77,KP91,Miller00}) with the main emphasis  given 
to the description of the nuclear structure in relativistic kinematics.

In the current work we focus on light-front treatment of  the electron-bound nucleon interaction.
Based on the effective light-front perturbation theory, we calculate the cross section of electron-bound nucleon scattering  
by explicit separation of the propagating (prop) and instantaneous (inst) contributions. Within light-front approach  we introduced 
a new,  parameter, $\eta$ which allows to quantify the off-shellness of the $\gamma^*N_{bound}$ scattering,  universally for any kinematical 
situation.  The derived expressions are compared with the off-shell cross sections which are 
currently being used in the description of electro-nuclear reactions.  We also present numerical analysis of our calculations where we identify 
kinematics in which off-shell effects can be suppressed or isolated for dedicated investigation of bound nucleon properties.  The numerical analysis 
allows us to conclude that by restricting the new off-shell parameter $\eta<0.1$  one can  confine the off-shell effects below $5\%$ for any realistic values of 
bound nucleon momenta at different $Q^2$ of electroproduction reaction.
 
 In Sec.\ref{II} we set up the
 calculations isolating the electromagnetic hadronic tensor for exclusive $d(e,e^\prime N)N$ scattering within 
 plane wave impulse approximation~(PWIA).  We discuss  here the main problems associated with 
 probing deeply bound nucleons, 
 namely the increased role of the vacuum-fluctuations and identification of the nuclear wave function for a  bound nucleon.
 Sec.\ref{III} presents the calculation of the PWIA diagram within effective light-front perturbation theory and identification of 
the propagating  and instantaneous 
 components of the bound nucleon electromagnetic current. We also introduce the boost invariant off-shell  parameter that 
 naturally quantifies the off-shell effects in the light-front approach.
 In Sec.\ref{IV} we present the results in the form of the 
 electron-bound-nucleon cross section $\sigma_{eN}$ which is compared with the predictions of other approaches in Sec.\ref{V}.
 Sec.\ref{VI} presents the summary of the results and outlook on possible extension beyond PWIA approximation.  
 In Appendix~\ref{App.A} we  give the diagrammatic rules of effective  light-front perturbation theory. The details of derivation of the bound nucleon 
 structure  functions are  presented in Appendix~\ref{App.B}.

\section{Setting-up to the Calculation}
\label{II}
The simplest case of electroproduction process
involving electron scattering from  a
bound nucleon is the reaction:
\begin{equation}
e + d \rightarrow e^\prime + N_f + N_r,
  \label{reaction}  
\end{equation}
in which one of the nucleons is knocked-out ($N_f$) by the virtual photon  and the other is
treated as a recoil ($N_r$).   Deuteron represents  as testing ground for development of many relativistic 
approaches in description of electro-nuclear processes (see e.g. \cite{FS77,Gross79,Glazek84,Miller13} ) 
which in principle  can be  generalized for medium to heavy nuclei.

\begin{figure}[ht]
  	\centering
 	\includegraphics[scale=0.5]{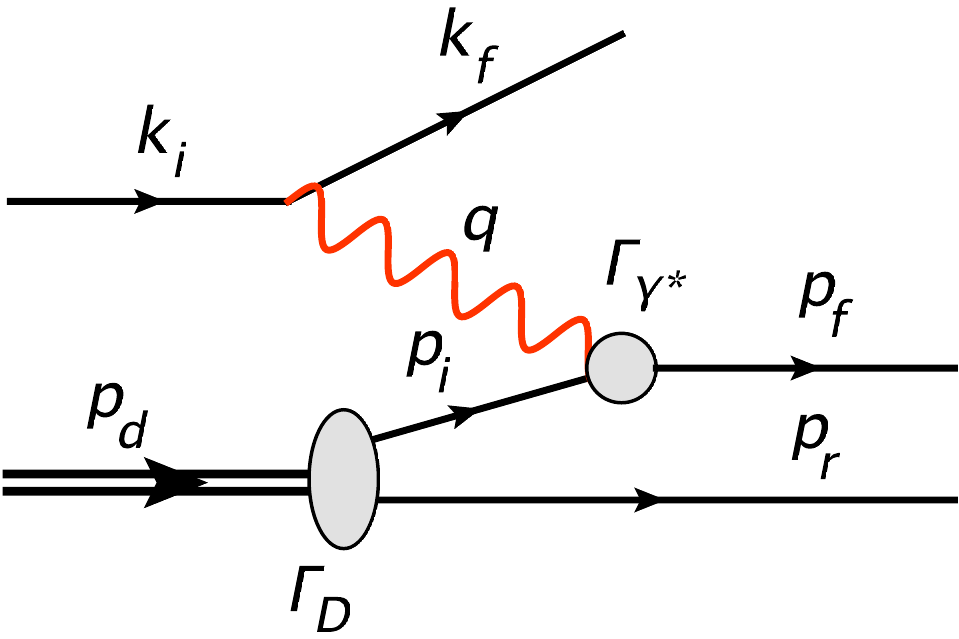}
	  \vskip -0.5cm
 	\caption{Exclusive electro-disinegration of the deuteron in plane wave impulse approximation:}
 	\label{Deuteron-PWIA}
  \end{figure}

For our purpose of  defining the cross section of electron-bound nucleon scattering we consider the single photon exchange case of the above 
reaction within  plane wave impulse approximation~(PWIA)  corresponding  to the diagram of Fig.\ref{Deuteron-PWIA}.  

Here, within PWIA the off-shellness of the bound nucleon is completely defined by the four-momentum of  the deuteron, $p_d$ and  
spectator nucleon $p_r$:  $p_i = p_D - p_r$.
The one photon-exchange approximation allows to factorize electron and hadronic parts of the interaction in the  invariant Feynman amplitude 
presented as follows: 
 \begin{equation}
{ \cal M} =    \langle \lambda_f\mid j_e^\nu\mid \lambda_i\rangle {e^2g_{\nu\mu}\over q^2 }  \langle s_f,s_r \mid A_0^\mu\mid s_d\rangle,
\label{M}
\end{equation}
where $q^2$ is the virtual photon's momentum squared. Here the leptonic current $j_e^\nu$ is defined as:
\begin{equation}
 \langle \lambda_f\mid j_e^\nu\mid \lambda_i\rangle =  \bar{u}(k_f,\lambda_f)\gamma^\nu u(k_i,\lambda_i),
 \end{equation}
 where  $\langle s_f,s_r \mid A_0^\mu\mid s_d\rangle$ represents  the invariant amplitude  of $\gamma^*d\rightarrow NN$ scattering, 
  
 Using Eq.(\ref{M}) for the differential cross section of reaction (\ref{reaction}) one obtains:
\begin{equation}\label{cross-section}
{d\sigma\over d^3k_f/\epsilon_f d^3p_f/E_f} = {1\over 4 \sqrt{(p_d \cdot k_i)^2}}  {e^4\over q^4} L^{\mu\nu} H_{\mu\nu} {\delta((q+p_d - p_f)^2 - m_N^2) \over 4 (2\pi)^5}.
\end{equation}
where, terms proportional to electron's mass squared ($m_e^2$) are neglected.  Here the leptonic tensor is defined as:
\begin{equation}\label{L-Tensor}
 L^{\mu\nu} = {1\over 2} \sum\limits_{\lambda_1 \lambda_2} \left( \bar{u}(k_f,\lambda_f)\gamma^\nu u(k_i,\lambda_i)\right)^{\dagger}  \bar{u}(k_f,\lambda_f)\gamma^\mu u(k_i,\lambda_i) 
\end{equation}
whereas the nuclear electromagnetic tensor is expressed through the scattering amplitude $A_0^\mu$ as follows:
\begin{equation}\label{H-Tensor}
H^{\mu\nu} = {1\over 3}   \sum\limits_{s_d s_r s_f} \langle s_d \mid A_0^{\mu \dagger} \mid s_f,s_r \rangle \langle s_f,s_r \mid A_0^\nu\mid s_d\rangle.
\end{equation}

 If one introduces  $\Gamma_{\gamma^*}^\mu$ and  $\Gamma_{D}$  invariant vertices (Fig.\ref{Deuteron-PWIA})  then
 within PWIA the amplitude $A_0^\mu$ can be   presented in the form:  
 \begin{equation}\label{A0_cov}
\langle s_f,s_r \mid A_0^\mu\mid s_d\rangle = -\bar u(p_f,s_f)\Gamma_{\gamma^*}^\mu \frac{\sh p_i + m_N}{ p_i^2-m^2_N} 
\cdot \bar u(p_r,s_r) \Gamma_{D}\cdot \chi^{s_d},
\end{equation}
where  $\chi^{s_d}$ is the  spin wave function of the deuteron. 

\begin{figure}[h]
  	\centering
 	\includegraphics[scale=0.9]{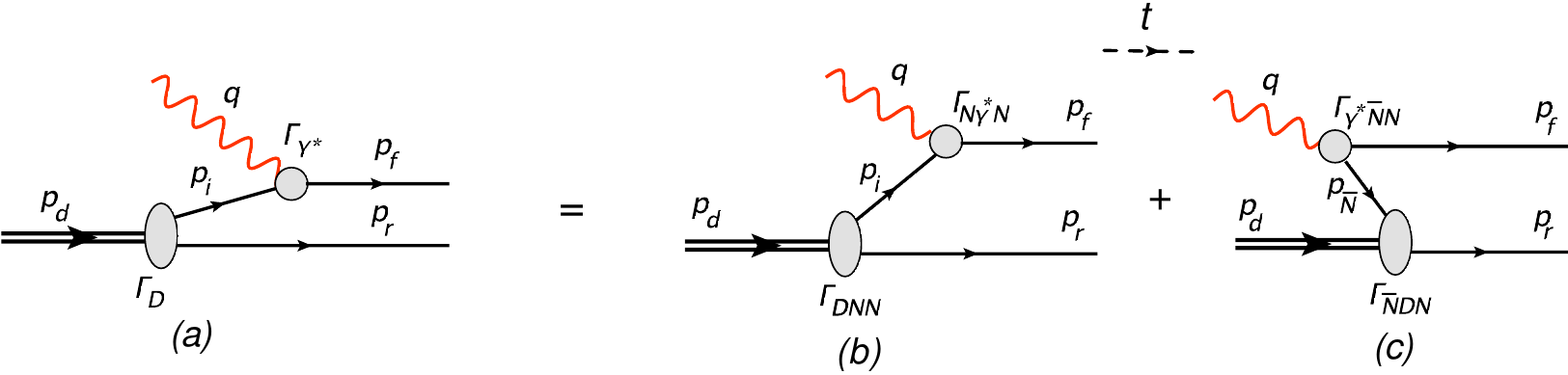}
	  \vskip -0.5cm
 	\caption{Representation of the covariant scattering amplitude (a) as a sum  two time-ordered 
	diagrams.  (a) Virtual photon scattering from the bound nucleon; (b) Production of the $\bar N N$ pair by the virtual photon 
	with subsequent absorption of the antinucleon  by the deuteron.}
 	\label{tordered}
  \end{figure}

As it follows from the above equation $A_0^\mu$ contains neither the electron-bound nucleon scattering nor the
nuclear wave function 
in the explicit form.  The $eN_{bound}$ scattering and the nuclear wave function  appears only when one considers the amplitude of Fig.\ref{Deuteron-PWIA} in a time ordered perturbation theory in which case 
the invariant Feynman diagram splits into  two time orderings as presented in Fig.\ref{tordered}.
  
Here, Fig.\ref{tordered}(b)  represents a
scenario in which
the virtual photon interacts with the preexisting  bound nucleon in the deuteron  with 
$\Gamma_{DNN}$ representing the vertex of $D\to NN$ transition and $\Gamma_{N\gamma^*N}$ 
the $\gamma^*N\rightarrow N$ electromagnetic interaction.
Fig.\ref{tordered}(c) however represents a
very different scenario, in this case the
virtual photon produces an
intermediate $\bar NN$ state  at  
the  $\Gamma_{\gamma^*\bar N N}$ vertex with subsequent absorption of the antinucleon, $\bar N$ in the deuteron at the $\Gamma_{\bar NDN}$ 
vertex.  The latter is not related to the $\gamma^*N$ scattering and the nucleon wave function in the deuteron.   Fig.\ref{tordered}(c) is commonly  
refereed as a "Z-graph" and is a
purely relativistic effect. As a result in the
non-relativistic limit one deals with the diagram of Fig.\ref{tordered}(b) 
only, which allows to express the covariant scattering amplitude through the nonrelativistic nuclear wave function and $\gamma ^* N_{bound}$ 
scattering amplitude.  However the situation  becomes complicated when one is interested in the bound nucleon momentum $p_i \sim M_N$, which 
can be probed at momentum transfer $q\gg M_N$.  In this case  
the
"Z-graph" contribution (Fig.\ref{tordered}(c)) becomes comparable with 
the one in Fig.\ref{tordered}(b) preventing the straightforward introduction of the
nuclear wave function. 

This situation is reminiscent to 
the QCD processes in probing partonic  structure of hadrons in which case due to the
relativistic nature of partons,  vacuum diagrams can 
not be neglected in the time ordered perturbation theory defined in the lab frame of the hadron\cite{Feynman}.   
The solution in this case  is  to consider
the scattering process in the infinite momentum frame (or on the light-front), which allows to suppress 
the "Z-graphs" and consider only the diagrams similar to Fig.\ref{tordered}~(b) for which one can introduce 
the wave function of the
constituents.

Our approach in probing deeply bound nucleon is similar to that of  the partonic model,  
in which  we consider the reaction~(\ref{reaction})  on
the light-front  allowing us to exclude the contribution of the vacuum diagrams (Fig.\ref{tordered}(c)) and introduce  a  
light-front nuclear wave function.    


\section{Derivation within Effective Light-Front Perturbation Theory}
\label{III}

\subsection{Scattering amplitude in PWIA}

\begin{figure}[ht]
  	\centering
 	\includegraphics[scale=0.9]{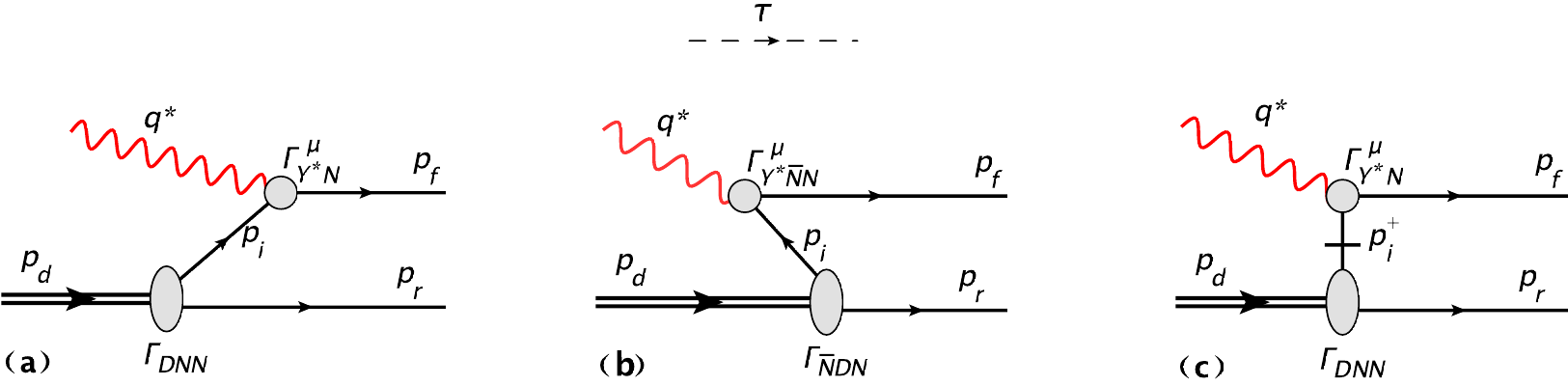}
	  \vskip -0.5cm
 	\caption{Representation of the covariant scattering amplitude  as a sum  of two light-cone ($\tau$)-time -ordered 
	diagrams as well as instantaneous interaction.  (a) Virtual photon scattering from the bound nucleon; (b) Production of the $\bar N N$ 
	pair by the virtual photon with subsequent absorption of the antinucleon  by the deuteron., (c) Instantaneous interaction of virtual photon with the bound nucleon.}
 	\label{tauordered}
  \end{figure}

We consider now the reaction (\ref{reaction})  on  the light-front, where the light-cone  time  is 
defined as  $\tau\equiv t+z$. 
To  calculate the PWIA amplitude of the  reaction (\ref{reaction}) we apply effective 
light-front    perturbation theory~(LFPT) in the  $\tau$-time ordered representation of the scattering amplitude $A_0^\mu$.  
In such approach the scattering amplitude (\ref{A0_cov}) is expressed as a sum of the noncovariant  diagrams presented 
in Fig.\ref{tauordered}. Here in addition to the two $\tau$ orderings analogous to the time ordering of Fig.\ref{tordered} one has 
an  additional contribution of Fig.\ref{tauordered}(c) corresponding to the instantaneous interaction due to
the spinor nature of the bound nucleon.

To proceed with the calculations we  choose  a
reference frame with  $z$ axis antiparallel to the transferred
momentum, $\hat z \ ||  \  -{\bf q}$,  such that the deuteron is aligned along $\hat z$.

The calculations are  performed in the Light-Cone (LC)  reference frame in which case
  where   the four-momenta are defined as 
$(p^+, p^-, p_x,p_y)$,  where $p^\pm = E\pm p_z$ with $p^+$ representing the light-cone longitudinal momentum.
 We employ  the $\gamma$-matrix algebra using the following light-cone definitions:
\begin{equation}\label{LC-gammas}
\gamma^\mu = (\gamma^+, \gamma^-, \gamma_1,\gamma_2),   \ \ \ \mbox{where}  \ \ \  \gamma^\pm = \gamma_0\pm \gamma_3.
\end{equation}
The scalar products and other properties of four-vectors as well as $\gamma$ matrices  on the light-front are given 
in Appendix~\ref{App.A}.

We also define the light-cone momentum fractions, which are Lorentz invariant quantities with respect to  
boosts along the $z$ direction:
\begin{equation}
\alpha_r = {2 p_r^+\over p_d^+}, \ \ \  \alpha_q = {2 q^+\over p_d^+}, \ \ \ \alpha_f = {2 p_f^+\over p_d^+},  \ \ \ \alpha_N = \alpha_f - \alpha_q = 2-\alpha_r.
\label{alphas}
\end{equation}
Here $\alpha_r$, $\alpha_q$ and $\alpha_f$ correspond to the fraction of LC 
"+" component of the deuteron momentum  carried by the
recoil nucleon, virtual photon and final knock-out nucleon respectively.
The light-cone momentum fraction of the bound nucleon $\alpha_N$ is defined through 
the energy-momentum conservation.

We proceed with  the calculation of the scattering amplitude corresponding to the diagrams of Fig.\ref{tauordered} by applying the 
Light-Front perturbation rules (\cite{KS(1970), LB(1980)}) in an effective theory in which one identifies effective vertices for
nuclear transition and electron-bound nucleon scattering (see Appendix~\ref{App.A}).  At each  vertices 
the transverse, $p_\perp$ and plus, $p^+$ components of momenta are conserved.  Because of the latter and the 
chosen reference frame in which $q^+ = q_0 - |\textbf{q}| < 0$ the diagram of the Fig.\ref{tauordered}(b) will not contribute since
the production of $\bar N N$ pair will require $q^+ >0$. 
The remaining diagrams represent the amplitudes in which the virtual photon knocks-out a bound nucleon which propagates from 
the $d\rightarrow NN$ transition vertex to the $\gamma^*N$ interaction point, $A^\mu_{prop}$  (Fig.\ref{tauordered}(a)), and the   instantaneous amplitude,  $A^\mu_{inst}$  (Fig.\ref{tauordered}(c)) in which  $d\rightarrow NN$ transition and $\gamma^*N$ interaction take place at the same light cone time -$\tau$. 
In both diagrams the  nucleus exposes its constituents and the scattering takes place off the bound nucleon, which allow us to introduce the light-front
nuclear wave function and the amplitude of $\gamma^*N_{bound}$ scattering.

We now apply the light-front diagrammatic rules\cite{LB(1980)} (summarized in Appendix~\ref{App.A}) which yields for the propagating 
part of the scattering amplitude (Fig.\ref{tauordered}(a)):
\begin{equation}
 \langle s_f,s_r \mid A_{prop}^\mu\mid s_d\rangle =  - \bar u(p_f,s_f)\Gamma_{\gamma^* N}^\mu 
  \  \frac{1}{p_i^+} \frac{(\sh p_i + m_N)_{on}}{ (p_d^- - p_r^- - p_{i,on}^- )} \bar u(p_r,s_r) \Gamma_{DNN} \chi^{s_d},
 \label{A0prop}
 \end{equation}
 where $p_d^-$, $p_r^-$ and  $p_{i,on}^-$
 are defined from the on-energy shell condition:  $p^- = {m_j^2 + p_{\perp,j}^2\over p_j^+}$ with $j = d, r, (i,on)$. 
The "on"  subscript in $(\sh p_i + m_N)_{on}$ indicates that  all the  components of the 
 bound nucleon light-cone momenta are taken on-energy shell. 
 
 For the instantaneous diagram of Fig.\ref{tauordered}(c) applying the rules of Appendix~\ref{App.A} one obtains:
 \begin{equation}
 \langle s_f,s_r \mid A_{inst}^\mu\mid s_d\rangle =  - \bar u(p_f,s_f)\Gamma_{\gamma^* N}^\mu   \frac{1}{p_i^+}\left(  \frac{1}{2}\gamma^+ \right)  
 \bar u(p_r,s_r) \Gamma_{DNN} \chi^{s_d}.
 \label{A0inst}
 \end{equation}
 Note that in 
 both expressions (\ref{A0prop}) and (\ref{A0inst}) one has the same nuclear, $\Gamma_{DNN}$ and electromagnetic, 
 $\Gamma_{\gamma^* N}$ vertices. 
 
 For further elaborations, we introduce 
 the off-energy shell "-" component of the
 bound nucleon $p_i^- = p_d^- - p_r^-$ ,
 and using the
 definition:
 $p_j^- = {m_j^2 + p_{j,\perp^2} \over p_j^+}$ for the on-energy-shell "-"  component for $ j = d,r$ 
  as well as Eq.(\ref{alphas}) one obtains:
 \begin{equation}
 \frac{1}{ p_d^- - p^-_r - p^-_{i,on} }  = \frac{1}{ p_i^- - p^-_{i,on} } = \frac{p_d^+} {M_d^2 - 4 \frac{(m_N^2 + p_\bot^2)}{\alpha(2-\alpha)} }.
\label{denom}
\end{equation}
Using the above relation as well as the sum rule relation for on-shell spinors:
\begin{equation}
(\sh{p}_{i} + m_N )_{on}= \sum_{s_i} \left( u(p_i,s_i) \bar{u}(p_i,s_i)\right)_{on},
\label{sumrule}
\end{equation} 
for the sum of the two amplitudes  in Eqs.(\ref{A0prop}) and (\ref{A0inst}) one obtains:
\begin{eqnarray}
A^{\mu} = A^\mu_{prop} + A^\mu_{inst} = && - \bar u(p_f,s_f)\Gamma_{\gamma^* N}^\mu \frac{\sum_{s_i} u(p_i,s_i) \bar{u}(p_i,s_i)}
{ \frac {\alpha}{2}\left( M^{2}_{d}-4\frac {m^{2}_N+{\bf p_T}^2}{\alpha \left( 2-\alpha \right) }\right)}  \bar u(p_r,s_r) \Gamma_{DNN}  \chi^{s_d}  \\
&&-  \bar u(p_f,s_f)\Gamma_{\gamma^* N}^\mu \frac{ \frac{1}{2} \gamma^{+} \left({p}_{i }^{-}  - {p}_{i,on}^{-} \right)}{  \frac {\alpha}{2}\left( M^{2}_{d}-4\frac {m^{2}_N+{\bf p_T}^2}{\alpha \left( 2-\alpha \right) }\right)} 
 \bar u(p_r,s_r) \Gamma_{DNN} \chi^{s_d}. 
\label{A0s}
\end{eqnarray}

Within PWIA we  can factorize the above expression in the form of 
a product of 
electromagnetic current and the
light-front nuclear wave function.  For this we introduce the
light-front wave functions in the form\cite{FS81,Artiles_Sargsian-multisrc1}:
\begin{equation}
\psi ^{s_{i}s_{r}s_{d}}_{LF }\left( \alpha ,{\bf p_T}\right) =-\frac {\bar {u}( p_{i}, s_i) \bar u( p_{r}, s_r) \Gamma _{DNN}\chi^{s_d}}{\frac {1}{2}\left( M^{2}_{d}-4\frac {m^{2}_N+{\bf p_T}^2}{\alpha \left( 2-\alpha \right) }\right) }\dfrac {1}{\sqrt {2\left( 2\pi \right) ^{3}}}.
\label{wflf}
\end{equation}
From the above definition one also obtains:
\begin{equation}
\frac{ 1}{\frac {\alpha}{2}\left( M^{2}_{d}-4\frac {m^{2}_N+{\bf p_T}^2}{\alpha \left( 2-\alpha \right) }\right)} \bar u(p_r ,s_r) \Gamma_{DNN} \ \chi^{s_d}  = - \sum_{s_i} \frac{ 	 u(p_i,s_i) }{2m_N}  \frac {\psi ^{s_{i} s_r s_d}_{LF}\left( \alpha, {\bf p_T}\right) }{\alpha }\sqrt {2\left( 2\pi \right) ^{3}}
\label{wfrel}
\end{equation}
Using now the above Eqs.(\ref{wflf}) and (\ref{wfrel}) in $A_{prop}^\mu$ and $A_{inst}^\mu$ respectively,  the Eq.(\ref{A0s}) can be presented 
in the form:
\begin{equation} 
A^{\mu }=  A^{\mu}_{prop} + A^\mu_{inst}  =  \sum _{s_i}J^{\mu }_{N}  \left( p_{f}s_{f},p_{i}s_{i}\right)  \dfrac {\psi ^{s_{i} s_r s_d}_{LF}\left( \alpha, {\bf p_T}\right) }{\alpha }\sqrt {2\left( 2\pi \right) ^{3}},
\label{A0sb}
\end{equation}
where we  introduced the electromagnetic current 
of  the bound nucleon as follows:
\begin{equation}
J^{\mu}_{N} (p_f s_f,p_i s_i) = \bar{u}(p_f s_f) \Gamma_{\gamma\ast N}^{\mu}  u(p_i s_i)  +  
\bar{u}(p_f s_f) \Gamma_{\gamma\ast N}^{\mu} \frac{ \gamma^{+} \left({p}_{i }^{-}  - {p}_{i \ on}^{-} \right)}{4m_N}  u(p_i s_i).
\label{J_N}
\end{equation}
Here, the Dirac spinor of the initial nucleon $u(p_i,s_i)$ is defined  for
the on-shell momentum, $p_{i,on} = (p^-_{i,on},p_{i}^+, p_{i}^\perp)$.
Our main focus in the following sections will be the electromagnetic current of Eq.(\ref{J_N},
which characterizes the   interaction of 
the electron probe  with the bound nucleon.

\subsection{Propagating and Instantaneous Components of Electromagnetic Current}
To identify   the
propagating and instantaneous  parts of the electromagnetic current in Eq.(\ref{J_N}) we consider first  
the electromagnetic vertex $\Gamma_{\gamma\ast N}^{\mu}$.
Since the final state of the interacting nucleon is  on mass shell, and  only  the   positive light-front
energy projections enter in the amplitude, we are led to the half off-shell vertex function in the general form
(see e.g. \cite{Bincer60,Koch90,Koch96}:
\begin{equation}
\Gamma_{\gamma^{\ast} N}^{\mu} = \gamma^{\mu }F_{1}  + i \sigma^{\mu \nu} q_{\nu} F_{2}  {\kappa \over 2 m_N} + q^{\mu}  F_{3}, 
\label{eNvertex}
\end{equation}
where the form-factors $F_{1,2,3}=F_{1,2,3}(m_N^2,p_i^2,q^2)$ are functions of  Lorentz invariants constructed from the momenta 
of initial and final nucleons and momentum transfer $q$.   In general one expects $F_{1,2}(m_N^2,p_i^2,q^2)$ not to be identical 
with the corresponding on-shell nucleon form-factors ($F_{1,2}(m_N^2,m_N^2,q^2)$).  This difference is due 
to the  modification of  the internal structure  of  nucleons  in the nuclear medium.  Such modification,  in principle,  should 
originate  from the dynamics similar to the
one responsible for the medium modification of partonic distributions of  bound nucleon - commonly referred as EMC effect (\cite{EMC}).
This however is out of the scope of our discussion since we are interested only in 
 the effects related to the off-shellness of the interacting 
nucleon's electromagnetic current.  Thus, in the numerical estimates we will use unmodified nucleon form-factors measured for free nucleons.
Concerning to $F_{3}$, it does not contribute  to the cross section of the process due to the gauge invariance of the leptonic current: 
$q_\mu j_e^\mu = 0$.  However for consistency one can estimate the $F_{3}$ form-factor based on the fact that due to 
the conservation of the momentum sum rule in light-front approach the electromagnetic current of the bound-nucleon is conserved: 
\begin{equation}
q_\mu J_N^\mu=0.
\label{qcons}
\end{equation}
Using Eq.(\ref{J_N}) together with (\ref{eNvertex}) one obtains: $ F_{3}  =  F_{1} { \shl{q}   \over Q^2}$. Inserting the later into Eq.(\ref{eNvertex})
one can separate the
propagating and instantaneous parts of the electromagnetic vertex in the form
\begin{equation}
\Gamma_{\gamma^{\ast} N}^{(prop) \mu} = \gamma^{\mu }F_{1}  + i \sigma^{\mu \nu} q_{\nu} F_{2}  {\kappa \over 2 m_N}, 
 \label{Vertex-on}
\end{equation} 
and 
\begin{equation}
\Gamma_{\gamma^{\ast} N}^{(inst) \mu} 
=    \left( \gamma^{\mu }F_{1}  + i \sigma^{\mu \nu} q_{\nu} F_{2}  {\kappa \over 2 m_N} \right)  \frac{  \Delta\sh{p}_i}{2m_N} - F_1{q^{\mu} \over q^2} \sh{q} \Big(  {\bf 1} + \frac{  \Delta\sh{p}_i}{2m_N}  \Big),
\label{Vertex-off}
\end{equation} 
where, $\Delta{p}_i^\mu=p^\mu_{i } - p^\mu_{i, on}$ and  $2 \Delta\sh{p}_i=\gamma^{+} \left( p^-_i - p^-_{i, on }\right)$ since $\Delta p_i^+ = \Delta p_{i}^\perp = 0$. In the following derivations we will use the relation:
\begin{equation}
\Delta{p}_i^- = -q^- + (p_f^- - p_{i,on}^-) = \dfrac{Q^2}{q^+} -   \dfrac{m^{2}_N+{\bf p_{T}}^2 }{p_f^+ p_i^+}  q^+ =  \frac{1}{p_d^+} \left(M_d^2 - 4 \frac{(m_N^2 + {\bf p_{T}}^2)}{\alpha(2-\alpha)} \right), 
\label{off-shell_factor}
\end{equation} 
as well as:
\begin{equation}
2  \Delta{p}_i \cdot p_i =  \Delta{p}_i^- p_i^+ = p_i^2 - m_N^2,
\end{equation}
which allow to express the electromagnetic current in boost-invariant variables.

The separation of the electromagnetic vertex into propagating and instantaneous  parts in  Eqs.(\ref{Vertex-on}) and (\ref{Vertex-off} allows to separate 
the electromagnetic current in Eq.(\ref{J_N}) into corresponding  parts in the following form: 
\begin{equation}
J_{N}^\mu (p_f s_f,p_i s_i) = J^{\mu}_{prop} (p_f s_f,p_i s_i) + J^{\mu }_{inst}(p_f s_f,p_i s_i) 
\label{Jsum}
\end{equation}
where,
\begin{eqnarray}
J^{\mu}_{prop} (p_f s_f,p_i s_i) &= & \bar{u}(p_f s_f) \Gamma_{\gamma\ast N}^{(prop) \mu}  u(p_i s_i) \nonumber  \\
J^{\mu }_{inst}(p_f s_f,p_i s_i) &= & \bar{u}(p_f s_f) \Gamma_{\gamma\ast N}^{(inst)  \mu}   u(p_i s_i).
\label{J_on_off} 
\end{eqnarray}
It is worth mentioning that even though the propagating vertex in Eq.(\ref{Vertex-on}) has the same form as the free on-shell nucleon 
vertex the corresponding electromagnetic current $J^{\mu}_{prop}$ 
does not correspond to an on-shell scattering amplitude, since   
$q^\mu\ne p_f^\mu - p_{i,on}^\mu$.  Also, the current  conservation (Eq.(\ref{qcons})) is satisfied only for the sum of the propagating and 
instantaneous currents in Eq.(\ref{Jsum}).

\subsection{Off-Shell Parameter of $eN_{bound}$ Scattering}
While the off-shell effects in the propagating vertex of Eq.(\ref{Vertex-on}) are kinematical, due to the fact that $q^\mu\ne p_f^\mu - p_{i,on}^\mu$, the 
off-shell effects in the instantaneous vertex are dynamical.  The latter interaction arises exclusively due to the binding of the nucleon.  As it follows 
from Eq.(\ref{Vertex-off}) the strength of the instantaneous vertex is proportional  to the magnitude of the
factor $\Delta  p_i^-$  defined in Eq.(\ref{off-shell_factor}).
One can express  the
$\Delta  p_i^-$ factor through a boost invariant quantities by defining the light-front reference frame such that the four-momenta of the 
deuteron, $p_d^\mu$ and   momentum transfer $q^\mu$ are:
\begin{eqnarray}
p_d^\mu & = &  ({Q^2\over m_N}, {m_d^2  m_N\over Q^2},\bf{0_T}) \\ \nonumber
q^\mu  & = & (-{Q^2 x\over m_N (1 + \sqrt{1 + {4m_N^2x^2\over Q^2}} \ )}, {m_N\over x} ( 1 + \sqrt{1 + {4m_N^2x^2\over Q^2}} ),\bf{0_T}).
\end{eqnarray}
Using above definitions one  introduces the off-shell parameter $\eta$ such that, 
\begin{equation}
\Delta{p}_i^-  = -{m_N}\eta,
\end{equation}
where, 
\begin{equation}
\eta = {1\over Q^2} \left(4 \frac{(m_N^2 + {\bf p_{T}}^2)}{\alpha(2-\alpha)} - m_d^2 \right).
\label{eta}
\end{equation}
As it will be shown in the derivations bellow, the parameter $\eta$  provides the universal measure of the off-shell effect  which combines both 
the resolution of the probe through the  $Q^2$ and the binding effects of the nucleon through the light-cone variables, $\alpha$ and $\bf{p_T}$.

\section{ Electron-Nucleon Scattering Cross Section  }
\label{IV}
In many practical applications one needs to evaluate the electron-bound-nucleon cross section $\sigma_{eN}$ as it is defined in Ref.\cite{deForest(1983)}.
Such a cross section is calculated within PWIA in which case using Eq.(\ref{A0sb})  the nuclear electromagnetic tensor of Eq.(\ref{H-Tensor}) can be expressed as follows:
\begin{equation}\label{H=H_N*rho}
H^{\mu \nu}=H^{\mu \nu }_{N}(p_f,p_i) \ \rho_d \left( \alpha, {\bf p_T}\right) \ \dfrac {2-\alpha }{\alpha ^{2}} \ 2\left( 2\pi \right) ^{3},
\end{equation}
where spin averaged light cone density matrix of the deuteron $\rho_d(\alpha, p_T)$ and bound nucleon electromagnetic tensor 
$H^{\mu \nu }_{N}(p_f,p_i)$ are defined in the following forms:
\begin{equation}\label{LF-density}
\rho \left( \alpha, {\bf p_T}\right) =  \dfrac {1}{2s_d+1} \cdot \frac {1}{2} 
\sum _{s_{d},s_i,s_r }\dfrac {\left| \psi ^{s_{i}s_{r}s_{d}}_{LF}\left( \alpha,{\bf p_T}\right) \right| ^{2}}{2-\alpha}
\end{equation}
and
\begin{equation}\label{HN}
H^{\mu \nu }_{N}=\frac {1}{2}\sum_{s_{i}s_{f}=-1/2}^{1/2}  J_{N}^{\nu}\left( p_{f}s_{f}, p_{i}s_{i}\right)^{\dagger}  J^{\mu }_{N}\left( p_{f}s_{f}, p_{i}s_{i}\right).
\end{equation}

Inserting now Eq.(\ref{H=H_N*rho}) into Eq.(\ref{cross-section}) 
the Lorentz invariant cross section of the reaction (\ref{reaction}) can be presented as follows:
\begin{equation}\label{Inv-DCS}
{d\sigma\over d^3k_f/\epsilon_f d^3p_f/E_f}  =   {1\over 2 p_d \cdot k_i}  {\alpha^2_{EM}\over q^4} L_{\mu\nu} H^{\mu \nu }_{N}\rho \left( \alpha, {\bf p_T}\right) \dfrac {2-\alpha }{\alpha ^{2} } \delta\left( p_r^2 - m_N^2\right),
\end{equation}
where $\alpha_{EM}=e^2/(4\pi)$.   Introducing the light-front nuclear spectral function in the form:
\begin{equation}\label{Spectral function}
S_d^{LF}(\alpha, {\bf p_T}) = \rho_d \left( \alpha, {\bf p_T}\right) \dfrac {2-\alpha }{\alpha ^{2} } \delta\left( p_r^2 - m_N^2\right),
\end{equation}
similar to Ref.(\cite{deForest(1983)}) one can present the differential cross section as a product of $\sigma_{eN}$ and the spectral function 
as follows:
\begin{equation}\label{factorized_cross-section}
{d\sigma\over  d\epsilon_f d\Omega_{k_f}  d^3p_f}  =   \sigma_{eN} \ S_d^{LF}(\alpha, {\bf p_T}),
\end{equation}
were 
\begin{equation}\label{sigma_eN}
\sigma_{eN} =  {1\over 2 m_D \epsilon_i} \ {\epsilon_f \over E_f} \  {\alpha^2_{EM}\over q^4} \ L_{\mu\nu} H^{\mu \nu }_{N}.
\end{equation}
Here $\epsilon_i$, $\epsilon_f$  are initial and scattered electron energies. The $E_f$  represents the energy of 
the knock-out nucleon.  	
	
It is worth mentioning that the expression in Eq.(\ref{factorized_cross-section}) is universal for any nuclei in which case one needs to
replace the deuteron spectral function by  the  light-front spectral function  of the nucleus being considered.

\subsection{Structure Functions  of Bound Nucleon}

\begin{figure}[h]
  	\centering
 	\includegraphics[scale=0.5]{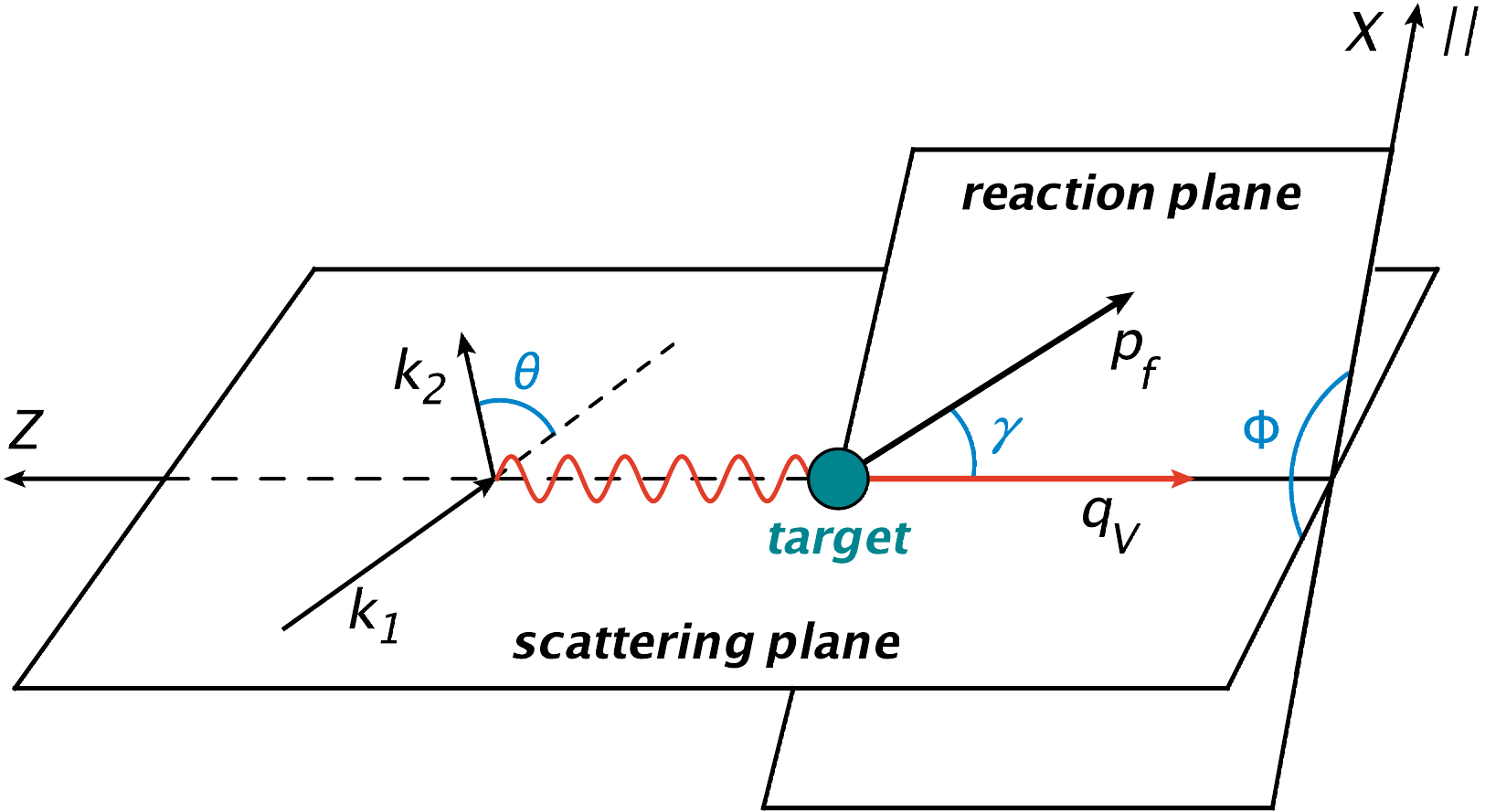}
	  \vskip -0.5cm
 	\caption{Definition of scattering and reaction planes of knock-out reaction.}
 	\label{sc_planes}
  \end{figure}

In calculating $\sigma_{eN}$ in Eq.(\ref{sigma_eN}) it is convenient to present it through the four independent structure functions 
of the nucleon $V^N_L$, $V^N_{TL}$, $V^N_T$ and $V^N_{TT}$ in the form:
\begin{equation}\label{sigma_eN-Gross}
\sigma_{eN} =  {1\over 2 m_D E_f}  \  {\sigma_{Mott}}  \left( \eta_L V_L^N + \eta_{TL} V_{TL}^N \cos\phi + \eta_{T} V_{T}^N + \eta_{TT} V_{TT}^N \cos(2\phi) \right),
\end{equation}
where $\sigma_{Mott} = {\alpha^2 \cos({\theta\over 2})^2\over 4 \epsilon_i^2 \sin({\theta\over 2})^4}$ with  $\theta$  being 
scattered electron angle. In the above equation:
 \begin{eqnarray}
\eta_L &=  & \frac{Q^4}{{\bf q}^4} \nonumber \\
\eta_T &=  & \frac{Q^2}{2{\bf q}^2}  +  \tan^2\frac{\theta}{2}  \nonumber  \\ 
\eta_{TT} &= &   \frac{ Q^2}{ 2 {\bf q}^2} \nonumber \\
\eta_{TL} &= &  \frac{Q^2}{  {\bf q}^2}\left(  \frac{Q^2}{{\bf q}^2}  +  \tan^2\frac{\theta}{2}  \right)^{1/2},
\end{eqnarray}
where  $Q^2 = 4\epsilon_i\epsilon_f\sin({\theta\over 2})^2$,  and $\bf q$ is the three momentum of the virtual photon.
The above defined  structure functions of the bound nucleon can be related to the light-front components of the 
nucleonic electromagnet tensor as follows (see Appendix \ref{App.B}):
\begin{eqnarray} \label{V-Gross}
	V^N_L & =& \frac{ {\bf q}^2 }{4 Q^2} \left( H^{++}{ Q^2 \over (q^+)^2 } + 2 H^{+-} + {(q^+)^2 \over  Q^2 } H^{- -} \right)\nonumber  \\
	V^N_{TL} & =& {|{\bf q}| \over q^+} \left( H^{+ \parallel}_N + H^{- \parallel }_N {(q^+)^2 \over Q^2} \right) \nonumber  \\ 
	V^N_T & =&  H^{\parallel \parallel }_N +  H^{\bot \bot} _N \nonumber  \\
		V^N_{TT}  & =& H_N^{\parallel \parallel} - H_N^{\bot \bot}   
\end{eqnarray} 
where $\pm$ correspond to $t\pm \hat z$ directions on the light-cone with $\hat z$ defined in the negative direction of 
of the transfered three momentum $\bf q$. The transverse components are chosen as follows: the perpendicular direction is defined by  $ \textbf{n}_{\bot}=\frac{\textbf{p}_f\times \textbf{q} }{|\textbf{p}_f\times \textbf{q}|}$, and the parallel unit vector projection is  $\textbf{n}_{\parallel}=\frac{ \textbf{q}  \times  \textbf{n}_{\bot} }{|\textbf{q}  \times  \textbf{n}_{\bot}|}$.  The scattering and reaction planes of the reaction are defined in Fig.\ref{sc_planes}.

Using now the Eq.(\ref{HN}) and the expression of the bound nucleon electromagnetic current from Eqs.(\ref{Jsum}), 
and (\ref{J_on_off}) one 
can calculate nucleon structure functions explicitly. 
In what follows we split the structure functions into two terms:
\begin{equation}
V^N_i  =  V^N_{i,prop} + V^N_{i,inst} \ \  \ \ \text{for i = L,TL,T,TT}
\end{equation}
where subscript "prop" corresponds to the structure functions calculated using  
the
propagating part of the  electromagnetic current, $J_{prop}^\mu$ only, 
while the terms with the subscript "inst" correspond to the contribution from $J_{inst}^\mu$ and its interference with $J_{prop}^\mu$.

Using the explicit forms of the   currents from  Eqs.(\ref{Jsum}),(\ref{J_on_off}) 
we calculate the above structure functions 
expressing them  through 
the
off-shell parameter $\eta$ (Eq.(\ref{eta}))  as follows\footnote{In  Appendix \ref{App.B} we also presented the same structure functions 
in more conventional form in  terms of scalar products of  kinematical variables  describing the reaction (see Eq.(\ref{V_LF})).}: 
\begin{eqnarray}
V_{L \ prop}^N  & = &   {\bf{q}}^2 \left[ F_1^2  \tau^{-1} \left(  1+ {p^2_{ \textbf{T}} \over m_N^2} + \tau \eta_i(\eta_i+\eta_q)   \right) -   F_1 F_2 {\kappa } \left( 2+ \eta_q   \right)   +  F_2^2 \kappa^2 \left(  {p^2_{ \textbf{T}} \over m_N^2} + \tau(1 + \eta_q) \right) \right],   \nonumber \\ 
\vspace*{0.3cm}
V_{L \ inst}^N & = & {\bf{q}}^2  \Big[ F_1^2  \eta_i \Big(  \tau \eta_i(1 + \eta_q)  - 2 - \eta_q    \Big)  + F_1 F_2 \kappa \Big( \tau \eta_i \left( 2-2\eta_i-\eta_q \right) +\eta_q \Big)    \nonumber \\
& & + F_2^2 \kappa^2 \tau  \Big( \tau \eta_i(\eta_i+\eta_q) - \eta_q   \Big) \Big],   \nonumber \\
\vspace*{0.3cm}
V_{TL \ prop}^N  & = &     2 \ |{\bf{q}}|  \  p_{\textbf{T}} \left(  F_1^2 +   F_2^2 \kappa^2  \tau  \right)\left[  2 +4{\alpha_N \over \alpha_q}  + 2\eta_i + \eta_q \right],  \nonumber \\ 
\vspace*{0.3cm}
V_{TL \ inst}^N  & = &   2 |{\bf{q}}|  p_\textbf{T}  \left(  F_1^2 +   F_2^2 \kappa^2  \tau  \right)  \left( 1 - \tau \eta_i   \right) \eta_q,  \nonumber 	\\
\vspace*{0.3cm}
V_{T \ prop \  \ }^N & =& \ 4m_N^2 \left[  F_1^2 \left(  {p^2_{ \textbf{T}} \over m_N^2} + 2 \tau(1 + \eta_q) \right) + 2  F_1 F_2 {\kappa } \tau \left( 2 + \eta_q   \right)  +  F_2^2 \kappa^2 \tau  \left(  2 + {p^2_{ \textbf{T}} \over m_N^2} + 2 \tau \eta_i(\eta_i+\eta_q)   \right) \right], \nonumber  \\
\vspace*{0.3cm}
V_{T \ inst \ \ }^N &  =&   2 Q^2 \Big[ F_1^2  \Big( \tau \eta_i(\eta_i+\eta_q) - \eta_q   \Big)  + F_1 F_2 \kappa \Big( \tau \eta_i \left( 2\eta_i+\eta_q - 2 \right) - \eta_q \Big)  \nonumber \\
& & + F_2^2 \kappa^2 \tau  \eta_i \Big(  \tau \eta_i(1 + \eta_q)  - 2 - \eta_q    \Big)  \Big],      \nonumber  \\
\vspace*{0.3cm}
V_{TT \ prop}^N  & = & 4 p_{\textbf{T}}^2 \left(  F_1^2 +   F_2^2 \kappa^2  \tau  \right), \nonumber \\ 
\vspace*{0.3cm}
V_{TT \ inst}^N  & =  &  0,   
\label{V_LFtoRP}
\end{eqnarray}
where, $\tau=Q^2/(4 m_N^2)$, $\ \eta_i=\eta \ \alpha_N/2$, $\ \eta_q=\eta \ \alpha_q/2$. Alternatively, one can write, 
\begin{eqnarray}
  \eta_i  & = &  - {2 \Delta{p}_i \cdot p_i  \over Q^2}=  {(m^{2}_N+{\bf p_{T}}^2 ) \over Q^2 } \dfrac{\alpha_q}{\alpha_f }  -  \dfrac{\alpha_N}{\alpha_q},  \\ 
 \eta_q   & = & -  {2 \Delta{p}_i \cdot q \over Q^2} =  {(m^{2}_N+{\bf p_{T}}^2 ) \over Q^2} \dfrac{\alpha_q^2}{\alpha_f \alpha_N}  -  1 .
\end{eqnarray}

The structure functions in Eq.(\ref{V_LFtoRP}) are Lorentz invariant and expressed through the boost invariant
variables $\eta$,  $\alpha_i$, $\alpha_q$ and $\alpha_f$. Since many experiments in  probing 
high momentum bound nucleons are  performed in the fixed target experiments  it is convenient to express the 
above variables  through the four momenta measured in the lab frame. Considering a Lab reference frame in which 
$\hat z || {\bf q}$, the  $\alpha_i$, $\alpha_q$ and $\alpha_f$ parameters can be expressed as follows:
\begin{equation}
\alpha_i = 2 - \alpha_r = \alpha_f - \alpha_q,  \ \ \  \alpha_r = {2 (E_r - p_r cos\theta_r) \over m_d}, \  \  \  
\alpha_q = {2 (q_0 - {\bf q})\over m_d}, \ \ \ \alpha_f = {2 (E_f - p_f cos\theta_f) \over m_d},
\end{equation}
where,  $p_d^\mu = (m_d,0)$, $q^\mu = (q_0, {\bf q})$, $p_r^\mu = (E_r, {\bf p_r})$ and  $p_f^\mu = (E_f, {\bf p_f})$   are four-momenta of 
the target deuteron, virtual photon, recoil and struck nucleon measured in the Lab frame.

\section{Numerical Estimates}
\label{V}

We present numerical estimates for kinematics  which will be explored in  experiments planned for 12 GeV upgraded Jefferson Lab.  In all calculations below we take the initial energy of the electron beam $\epsilon_i = 11$~GeV.

To quantify the extent of the binding effects we consider the ratio:
\begin{equation}
R = {\sigma_{eN}\over \sigma_{eN}^{on}},
\label{R}
\end{equation}
where $\sigma_{eN}$ is the cross section of electron bound nucleon scattering defined in Eq.(\ref{sigma_eN})  for 
given  initial momenta $\bf {p_i}$ or ($\alpha_i$ and $p_T$),  while 
$\sigma_{eN}^{on}$ corresponds to the same cross section for the electron scattering off the free moving 
nucleon with the same initial momenta. 

\begin{figure}[th]
\vskip 0.3cm
\centering
 	\includegraphics
	[height=7cm,width=16cm]{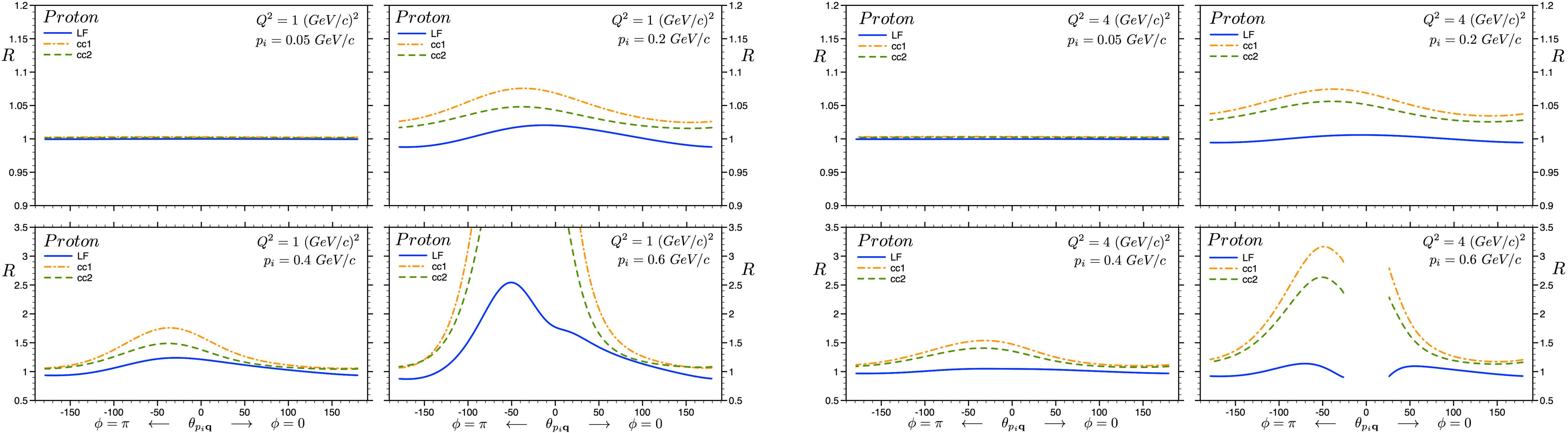}
	  \vskip -0.4cm
 	\caption{The $\theta_{p_iq}$ dependence of ratios of the off-shell cross section of electron-bound proton
	 scattering to that of 	the one-shell cross section. The solid lines are LF approximation, dashed and dash-dotted
	 curves corresponds to CC2 and CC1 versions of de Forest approximation\cite{deForest(1983)}. The panels 
	 correspond to the bound nucleon momenta $p_i=50, 200,400$ and $600$~MeV/c for 
	$Q^2 = 1$ and $4$~(GeV/c)$^2$.  The {\em minus}  sign of $\theta_{pq}$ indicates on kinematics corresponding 
	to $\phi=180^0$ between scattering and reaction planes. Calculations done for initial electron energy $\epsilon_i = 11$~GeV.}
 	\label{theta_dep_Q2_p}
  \end{figure}

First,  we consider the dependence of $R$ on "traditional" kinematical parameters  which define the 
electronuclear processes such as initial momentum of the bound nucleon ($p_i$) its relative angle with 
respect to  the transferred 3-momentum
($\bf q$) as well as
 the virtuality of the transferred momentum ($Q^2$).  
Additionally  we compare the predictions of LF approximation  with that of  the de Forest 
formalism\cite{deForest(1983)} which is commonly used  in the analysis of the experimental data. 
In all these estimates we use the same parameterization for the electric and magnetic form-factors of 
the nucleons. These parameterizations  are the same for the free nucleon. Thus we do not consider the 
effects related to the possible modification of the charge and magnetic current distributions in the bound nucleon.

In Fig.{\ref{theta_dep_Q2_p}  and Fig.\ref{theta_dep_Q2_n} we compare the angular dependences
of ratio  $R$  at different values of missing momenta at fixed $Q^2=1$ and $4$~(GeV/c)$^2$ for bound proton and 
neutron respectively.  As  Fig.{\ref{theta_dep_Q2_p} shows  LF approximation predicts 
off-shell effects for $Q^2=1$ (GeV/c)$^2$ as large as $40-250\%$  for bound proton momenta $\ge 400$~MeV/c.
Even larger effects are expected within the de Forest approach\cite{deForest(1983)}. We observe also that  the prediction within LF and de Forest approximations
significantly diverge close to the kinematical limit of  the scattering process as it can be seen in calculations for $p_i = 600$~MeV/c.

 \begin{figure}[th]
 \vskip 0.25cm
	\centering
 	\includegraphics
         [height=7cm,width=16cm]{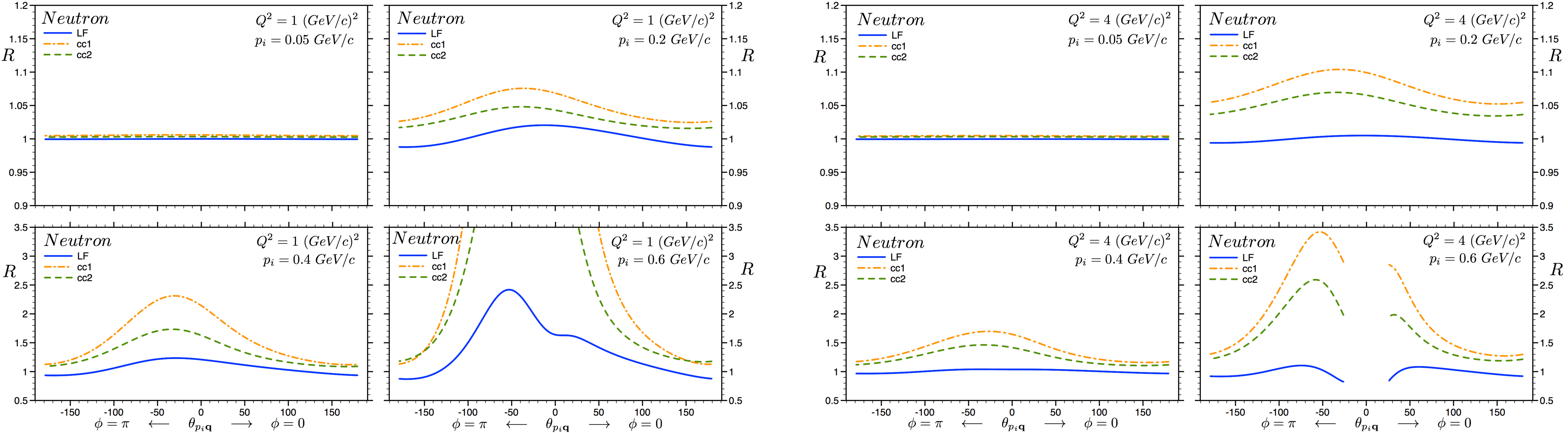}
	  \vskip -0.4cm
 	\caption{The same as in Fig.\ref{theta_dep_Q2_p}  but for scattering from bound neutron.}
 	\label{theta_dep_Q2_n}
  \end{figure}
 
Because of different  magnitudes and signs of  form-factors one predicts somewhat different off-shell 
effects  for scattering from a bound proton or neutron.   However, qualitatively the dependences of $R$ for 
kinematical parameters of the reaction for both proton and neutron are similar.

\begin{figure}[th]
  	\centering
 	\includegraphics[scale=0.3]{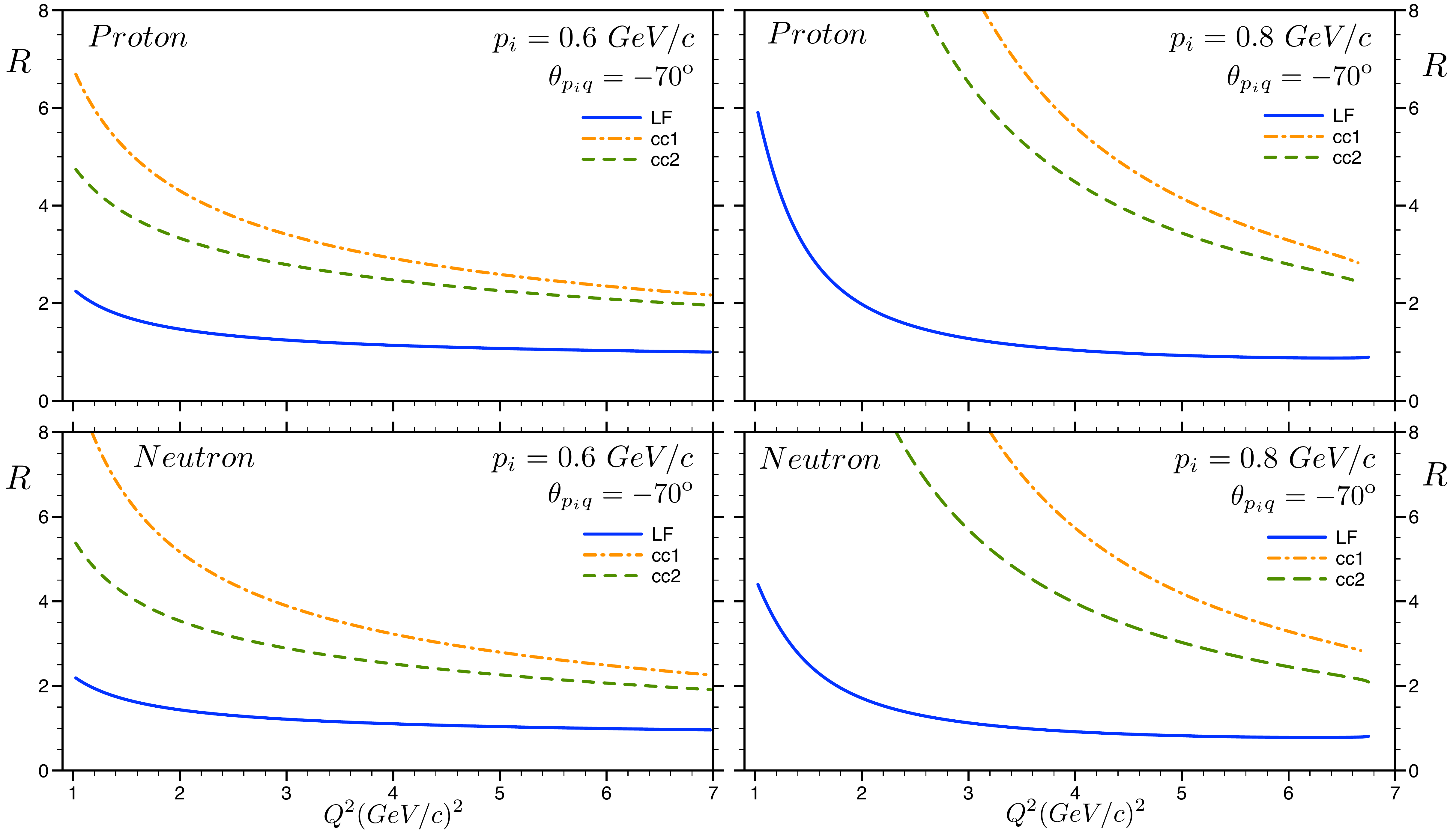}
	  \vskip -0.5cm
 	\caption{The $Q^2$ dependence of the off-shell effects for $\theta_{p_iq} = -70^0$ for proton and neutron targets.}
 	\label{Q2_dep_p_n}
  \end{figure}

An important feature of LF calculations following from Fig.\ref{theta_dep_Q2_p} and Fig.\ref{theta_dep_Q2_n} is the diminishing of  the off-shell effects with an increase of $Q^2$.
This reflects the dynamical nature of the LF approximation in which case the harder  the probe (larger $Q^2$) lesser is 
the sensitivity to the binding effects of the target nucleon. 
 It is worth mentioning that no such behavior  exists in the de Forest approximation since in this case part of the off-shell effects 
 are kinematical in which the energy of bound nucleon  is taken to be equal to the on-shell energy for the given momentum of the nucleon, with  
 the phase space of the initial nucleon being proportional to ${1\over \sqrt{m_N^2 + p_i^2}}$.

To ascertain the  extent of the $Q^2$ suppression on the  off-shell effects,  in Fig.\ref{Q2_dep_p_n} we present the $Q^2$ dependence of the ratio $R$ for proton and neutron initial momenta of 
$p_i = 600$~ and $800$~MeV/c.  Here we choose $\theta_{p_iq} = -70^0$ for which large off-shell effects 
are observed in Fig.\ref{theta_dep_Q2_p} and Fig.\ref{theta_dep_Q2_n}.  These calculations indicate that already 
at $Q^2\ge 4$~GeV$^2$ the  off-shell effects predicted in light-front approximation  are not more than $10\%$ for such a large 
bound nucleon momenta.

\begin{figure}[th]
  	\centering
 	\includegraphics[scale=0.32]{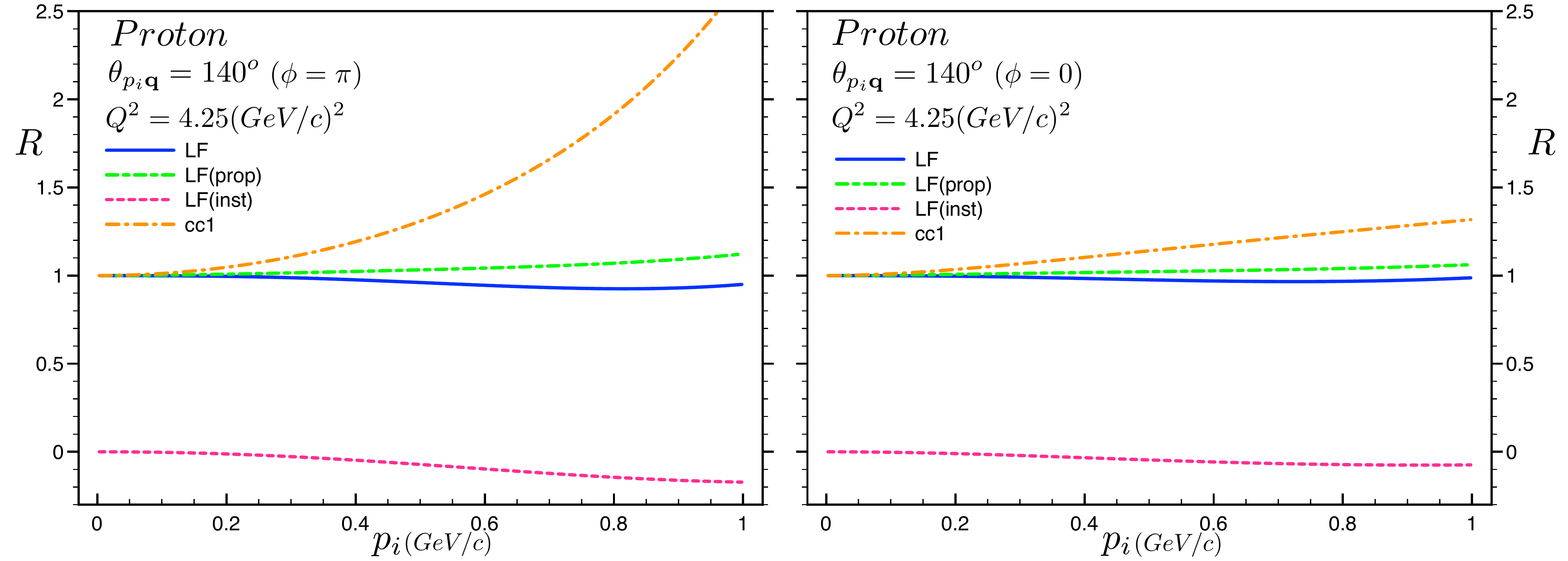}
	  \vskip -0.5cm
 	\caption{Off shell effects expected for the experiment of Ref.\cite{Boeglin:2014aca} (left panel).  The right-panel is the similar effects for $\phi = 0$ kinematics. }
 	\label{pi_dep_p}
  \end{figure}

For practical purposes in Fig.\ref{pi_dep_p} we estimate the dependence of the off-shell effects on  
the momentum of the bound nucleon for  kinematics relevant to the planned  JLAB experiment\cite{Boeglin:2014aca}   which is aimed at  probing deuteron structure at very large internal momenta.
As the figure shows for both cases  of  the angles  between scattering and reaction planes ($\phi$)  the  light-front approach  predicts off-shell effects to be less than $8\%$ for all kinematics with the latter value   happening at   $p_i = 850$~MeV/c.

At the end of the section we discuss whether the parameter $\eta$  introduced in Eq.(\ref{eta}) can be used as a 
universal parameter for estimation of the of-shell effects for any kinematic conditions of  electro-production reaction.
For this, in Fig. \ref{eta-graph} we calculate the $\eta$ dependence of $|R-1|$ for very large magnitudes  of bound nucleon 
momenta ($p_i = 600$ and $800$~MeV/c)  at different values of transverse momentum $p_T$.  Note that, the expected off-shell  effects
will be much less for smaller values of $p_i$.

  \begin{figure}[th]
 	\centering
 	\includegraphics[scale=0.3]{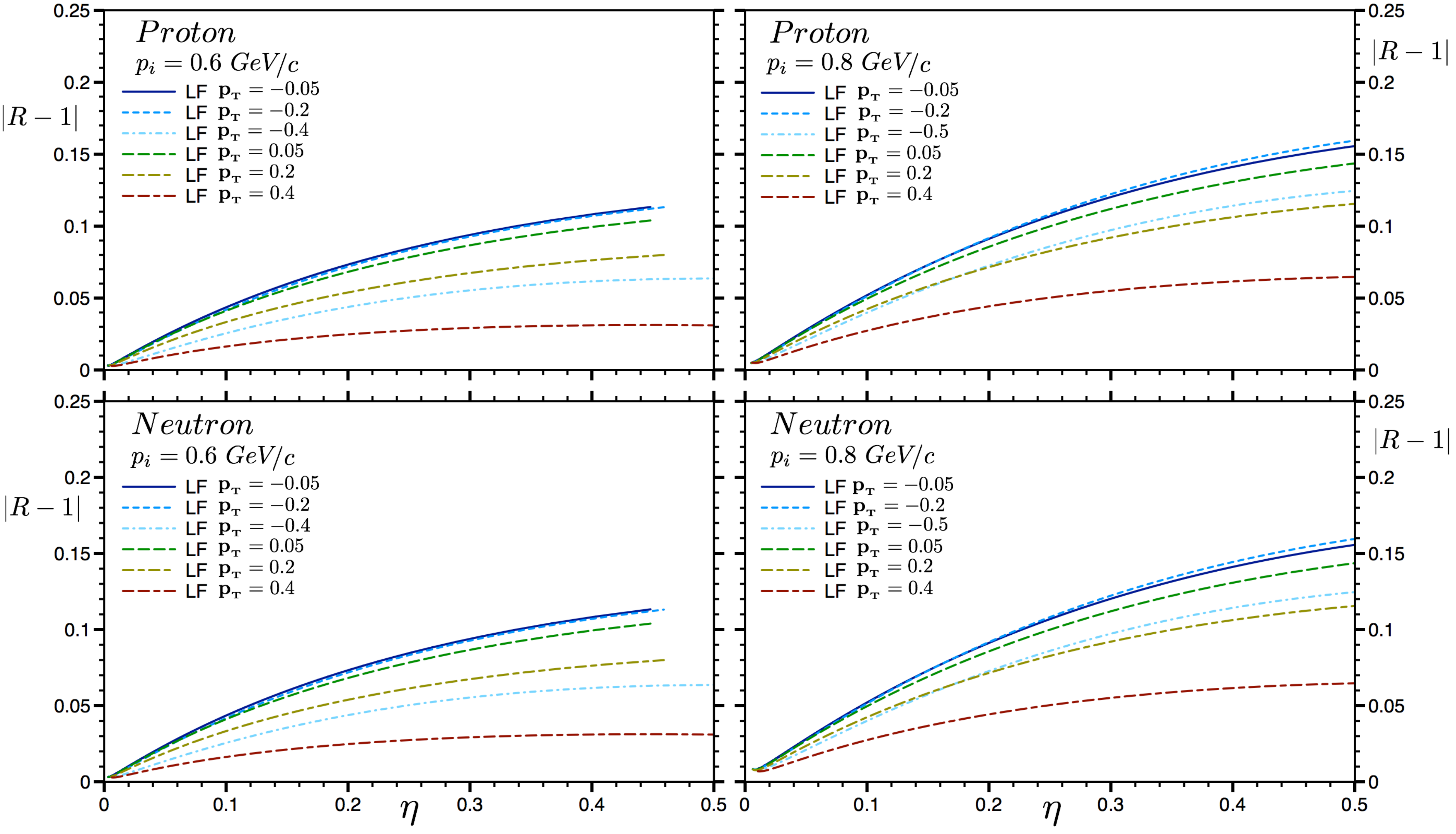}
 	\vskip -0.5cm
 	\caption{ The $\eta$ parameter dependence of the off-shell effects $|R-1|$ for $p_i = 0.6$ and $0.8$~GeV/c at different 
	values of the transverse momentum $p_T$. }
 	\label{eta-graph}
 \end{figure}
As the figure shows for any possible scenarios of  kinematics the off shell effects can be confined below $5\%$ as soon as 
$\eta <0.1$.  This represents a strong indication that the variable $\eta$ can be considered as an universal  parameter for  
controlling the off-shell effects in the reaction mechanism of electron-nuclear processes.

\section{Summary and Outlook}
\label{VI}

Based on  light-front  approach we calculated electron-deuteron scattering within PWIA which allowed us to 
isolate the electron-bound-nucleon scattering cross section, $\sigma_{eN}$.  Within LF approximation the vacuum 
contribution naturally disappears while the off-shell nature of the  nucleon 
results in a appearance  of an  instantaneous term in the electromagnetic current  of electron-bound nucleon scattering.
In deriving $\sigma_{eN}$ we separated the propagating and instantaneous contributions in the electromagnetic 
current which allowed explicitly to trace the effects associated with the binding of the nucleon.  

Furthermore in LF approach we were able to identify the parameter (defined as $\eta$) that universally characterizes the
extent of the off-shellness of 
electromagnetic current.

The derived $\sigma_{eN}$ is used to estimate the expected off-shell effects in electro-nuclear processes 
in kinematics relevant to the 12 GeV energy upgraded Jefferson Lab experiments.  We compared the 
LF predictions with that of the de Forest approximation widely used by experimentalists to estimate the 
off-shell effects in the reaction mechanism of electro-nuclear processes. These comparisons indicate that practically in 
all kinematic cases the  LF approach predicts less off-shell effects at $Q^2\ge 1$~GeV$^2$ than the de Forest
 approximation does. Most importantly the LF approach predicts a significant drop of the off-shell effects with an increase 
 of $Q^2$ which intuitively can be understood as  a decrease in the sensitivity 
 of the hard processes on the off-shellness 
 of the target nucleon.

We also checked our conjecture that the $\eta$-variable can be considered as a universal parameter in controlling 
off-shell effects. We found that for wide range of kinematics the off-shell effects can be suppressed on the level of 
$5\%$ as as soon as $\eta < 0.1$. The latter gives an effective method 
for  controlling the uncertainties 
in
the reaction mechanism for large varieties of electro-nuclear processes probing deeply bound nucleons in
 the nucleus.

Finally, it is worth mentioning that even though we  considered the $eA$  scattering  within  PWIA the 
obtained expressions for electromagnetic current are applicable also for scattering amplitudes in which 
the final state interaction between outgoing nucleons is considered within eikonal approximation.  
In this case (see e.g. Refs.\cite{gea,ms01}) the main part of the re-scattering amplitude is evaluated at the 
pole value of the struck nucleon propagator in the intermediate state.   As a result the entered electromagnetic 
current is again  half-off-shell  as the considered electromagnetic current  in  Eq.(\ref{Jsum}).

\acknowledgments
We are thankful to Dr. Werner Boeglin  and Chris Leon for numerous discussions on the physics of high energy electro-nuclear reactions. 
This work is supported by the US  DOE grant under contract DE-FG02-01ER-41172.
\appendix

\section{Light Front Perturbation Theory Rules}\label{App.A}

We present here a  summary of the rules for computation of  amplitudes within Light Front (LF) formalism. 

The LF scalar product and notations are (Lepage-Brodsky convention\cite{LB(1980),BPP(1997)}):
\begin{eqnarray}\label{x*p}
 &  & \ x \cdot p={ 1 \over 2} \left( x^+ p^-  + x^- p^+\right)  -  \textbf{x}_{\textbf{T}} \cdot  \textbf{p}_{\textbf{T}} \nonumber \\
 & & x^{\mu }=\left( x^{+}, {x}^-, x, y \right) = \left( x^{+}, {x}^-, \textbf{x}_{\textbf{T}}\right) \ , \  p^{\mu }=\left( p^{+},p^{-}, \textbf{p}_{\textbf{T}}\right) \nonumber  \\
 &  & x^{\pm} = t \pm z
\end{eqnarray} 
\vskip 0.3cm

\begin{figure}[h]
			\includegraphics[scale=0.7]{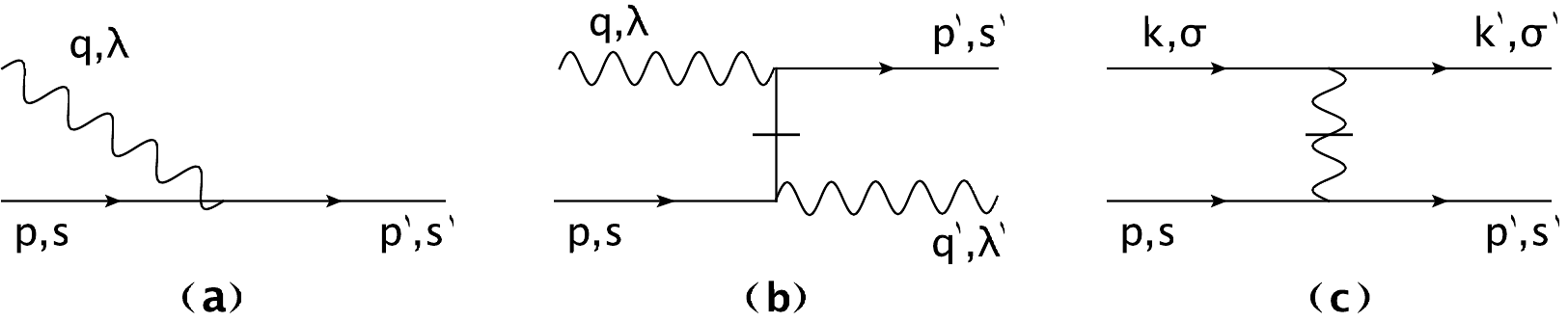}
	\centering
	\caption{Example of the scattering amplitude on the light-front ($\tau=x^+$ flows from left to right).}
	\label{LCPT-Vertices-QED}
\end{figure}

\vskip -0.3cm

Diagrammatic Rules for  effective  light-front perturbation theory can be formulated as follows:
\begin{enumerate}
	\item Draw all topologically distinct $\tau \equiv x^+$-ordered diagrams at the desired coupling power. 
	In addition to  the usual advanced and retarded propagation  between two events  one needs to include  a third possibility 
	in which the two events connected by an internal  fermion or photon interact   at the same LF $\tau$- time, commonly referred as 
	instantaneous term.
	\item Assign to each line a four-momentum $p^\mu$ and spin $s$ (or helicity, $\lambda$) corresponding to a single on-mass-shell particle, i.e. $p^2 = m^2$. 
	\item With spin 1/2 fermions associate on-mass-shell spinors $u(p, s)$, with antifermions $v(p, s)$, with photons $\epsilon_\mu(q,\lambda)$, etc, such that, 
	\begin{eqnarray}\label{Spinors-Norm}
	\bar{u}(p, s') u(p, s)  & =  & - \bar{v}(p, s') v(p, s) = 2m\delta_{s  s'} \nonumber \\
	\sum_{s}  u(p,s) \bar{u}(p,s) &  = &  \sh{p} + m \nonumber  \\  
	\sum_{s}  v(p,s) \bar{v}(p,s) & = &  \sh{p} - m   \nonumber \\
	\epsilon^\mu(q, \lambda') \epsilon^\mu(q, \lambda) & = & - \delta_{\lambda '  \lambda} \ , \ \  q \cdot \epsilon(q, \lambda) = 0  \nonumber \\
	\sum_{\lambda} \epsilon^\mu(q, \lambda) \epsilon^\nu(q, \lambda) & = &  - g^{\mu \nu } + { q^\mu \eta^\nu + q^\nu \eta^\mu \over q \cdot \eta}
	\end{eqnarray} 
	where $\eta$ is a null vector ($\eta^2=0$), given in LC gauge by,  $ \eta = (0,2,0,0)$
	
	\item  Each intermediate state gets a factor (inverse of the difference of the sums  over initial   and intermediate LF energies ($p_-$):
	{\small 
		\begin{equation}
		\frac{1}{ \sum_{ini} p^- - \sum_{int} p^-  + i\epsilon} 
		\end{equation}
	}
	where, '$ini$' stands for the initial state of the diagram and '$int$' for intermediate states. All particles are on-mass-shell, that is:  $\  p^- = \frac{ m^2 + p_{\textbf{T}}^2 }{p_i^+} > 0$.

	\item Internal lines account for two kind of interactions:
	\vskip .1cm
	\begin{itemize}
		\item Propagating, in which case, for a vertex like in Fig.(\ref{LCPT-Vertices-QED})(a) one has: 
		\begin{equation}
		\Gamma \ \bar{u}(p',s') \sh{\epsilon}(q,\lambda) u(p,s) \delta^2(\sum_{in} p_{\textbf{T}} - \sum_{out} p_{\textbf{T}} ) \delta(\sum_{in} p^+ - \sum_{out} p^+ ) 		\end{equation}
		where, '$in$'  and '$out $' mean flowing into and out of the vertex. 
		The $\delta$ functions at the vertex provide an explicit conservation of the plus and transverse components of '$in$' and '$out$' momenta.
		
		\item Instantaneous. For each vertex like  in Fig.(\ref{LCPT-Vertices-QED})(b)  (fermionic), include,
		\begin{equation}
		\Gamma^2 \ \bar{u}(p',s') \sh{\epsilon}(q',\lambda')  {\gamma^+ \over 2(q^+ - p'^+) } \sh{\epsilon}(q,\lambda) u(p,s) \delta^2(\sum_{in} p_{\textbf{T}} - \sum_{out} p_{\textbf{T}} ) \delta( \sum_{in} p^+ - \sum_{out} p^+ ).
		\end{equation}
		 And, for each vertex like in Fig.(\ref{LCPT-Vertices-QED}) (c)  (vector), include,
		\begin{equation}
		\Gamma^2 \ \bar{u}(p',s') \gamma^+ u(p,s)  {1 \over (p'^+ - p^+)^2 } \bar{u}(k',\sigma') \gamma^+ u(k,\sigma) \delta^2(\sum_{in} p_{\textbf{T}} - \sum_{out} p_{\textbf{T}} ) \delta( \sum_{in} p^+ - \sum_{out} p^+ ).
		\end{equation}
		
	\end{itemize}
	Here $\Gamma$ factors represent effective vertices which can be specified for the particular case of the 
	scattering.  They can correspond to electron-bound nucleon scattering as well as nuclear transition to the 
	constituent nucleons.  The conditions for for which such an effective vertices can be introduced in the Feynman diagrams 
	are discussed in Ref.\cite{ms01}.
	
	\item Sum over polarizations and integrate over each internal line with the factor,
	{\small$$ \sum_{s} \int  \frac{dp_{\textbf{T}} dp^+}{2(2\pi)^3 p^+} \Theta(p^+)   $$ }
	which ensures the plus component positivity (all particles move forward in LC time.)  
	
	\item Include symmetry factors. Also, a factor of -1 for each fermion loop, for  fermion lines beginning and ending at the initial state, and for each diagram in which fermion lines are interchanged in either of the initial or final states, as well as the overall sign from Wick's theorem. 
	
\end{enumerate}

\section{Nucleonic Tensor}\label{App.B}

Substituting Eq.(\ref{Jsum}) into Eq.(\ref{HN}), allows us to express  the nucleonic tensor as a sum of two terms:
\begin{equation}\label{HNsum}
H^{\mu \nu}_{N } = H^{\mu \nu}_{N \ prop} + H^{\mu \nu}_{N \ inst},
\end{equation}
where the the propagating contribution is given by,
\begin{equation}\label{HN-on_JJ}
H^{\mu \nu}_{N \ prop} =  {1 \over 2} \sum_{s_i s_f} (J^{s_i s_f \ \mu}_{prop} )^{\dagger}  (J^{s_i s_f}_{prop} )^\nu  = \frac{1}{2}Tr\left[\overline{\Gamma}_{\gamma^\ast N}^{(on) \mu} (\sh{p}_f + m_N) \Gamma_{\gamma^\ast N}^{(on) \nu} (\sh{p}_{i,on} + m_N) \right],
\end{equation}
and the instantaneous by,
\begin{eqnarray}\label{HN-off_JJ}
H^{\mu \nu}_{N \ inst} &=& {1 \over 2} \sum_{s_i s_f} \left( (J^{s_i s_f \ \nu}_{off})^{\dagger}  J^{s_i s_f \ \mu}_{inst}  + (J^{s_i s_f \ \nu}_{prop})^{\dagger}  J^{s_i s_f \ \mu}_{inst}  + (J^{s_i s_f \ \nu}_{inst})^{\dagger}  J^{s_i s_f \ \mu}_{prop}  \right) \\  \nonumber
&=&   \dfrac{1}{2}  Tr\Big[    \overline{\Gamma}_{\gamma^\ast N}^{(off) \nu} (\sh p_f + m_N) \Gamma_{\gamma^\ast N}^{(off) \mu} (\sh{p}_{i,on} + m_N)  \\ 
&&  +  \overline{\Gamma}_{\gamma^\ast N}^{(on) \nu} (\sh p_f + m_N) \Gamma_{\gamma^\ast N}^{(off) \mu} (\sh{p}_{i,on} + m_N)  +   \overline{\Gamma}_{\gamma^\ast N}^{(off) \nu} (\sh p_f + m_N) \Gamma_{\gamma^\ast N}^{(on) \mu} (\sh{p}_{i,on} + m_N)  \Big], \nonumber
\end{eqnarray}
where,  $\overline{\Gamma}_{\gamma^\ast N}^{\mu} = \gamma^0 \left( {\Gamma}_{\gamma^\ast N}^{\mu}\right)^{\dagger} \gamma^0 $. 
Notice that the initial momentum of the nucleon, $p_i$, occurring from now-on  corresponds to $p_{i,on} $, which allows to drop the on-shell label
"on"   without confusion. 
With this, we can write propagating  and instantaneous  contributions of the tensor, $H^{\mu,\nu}$  as functions of the nucleon form 
factors $F_1$ and $F_2$ as follows:
\begin{eqnarray}\label{HN-on}
H^{\mu \nu}_{N \ prop} & =& \ 2 F_1^2 \Big[ g^{\mu\nu} \left( m_N^2 - p_i \cdot p_f \right) + \left( p_i^{\mu} p_f^{\nu}  + p_i^{\nu} p_f^{\mu} \right)    \Big] 
\nonumber \\
& + &   F_1 F_2 \kappa  \Big[ 2 g^{\mu\nu} q \cdot \left( p_f - p_i\right) + \left( p_i^{\mu} q^{\nu}  + p_i^{\nu} q^{\mu} \right)  -  (  p_f^{\mu} q^{\nu}   +p_f^\nu q^{\mu} ) \Big] \nonumber \\
& +  & F_2^2  {\kappa^2 \over 2 m_N^2}   \Big[  g^{\mu\nu} \big[ q^2 \left( p_i \cdot p_f + m_N^2 \right)  - 2 \ q \cdot p_i \ q \cdot p_f \big] - q^2 \left( p_i^{\mu} p_f^{\nu}  + p_i^{\nu} p_f^{\mu} \right) \nonumber  \\
&  & \ \ \ \   -  q^{\mu}q^{\nu}  \left( p_i \cdot p_f + m_N^2 \right)  +  q \cdot p_f \left( p_i^{\mu} q^{\nu}  + p_i^{\nu} q^{\mu} \right) +  q \cdot p_i \left( p_f^{\mu} q^{\nu}  + p_f^{\nu} q^{\mu} \right),  \Big] 
\end{eqnarray}
and the instantaneous correction as follows:
\begin{eqnarray}\label{HN-off}
& & H^{\mu \nu}_{N \ inst} =   \ 2 F_1^2 \Big[ g^{\mu\nu} \Big(  \Delta p_i \cdot\left( p_i -  p_f \right) - { \Delta p_i \cdot p_i \over  m_N^2}  \Delta p_i \cdot p_f  \Big)  + \left(\Delta p_i^\mu p_f^{\nu}  + \Delta p_i^{\nu} p_f^{\mu} \right) \Big( 1 + {\Delta p_i \cdot p_i \over  m_N^2 }   \Big)  \nonumber \\
& +  & {2 \over q^2} q^{\mu}q^{\nu}  \Big( {2 \over q^2} q \cdot p_{f} \ q \cdot ( \Delta p_i + p_{i} ) -   ( p_{i} - p_f) \cdot ( \Delta p_i + p_{i} ) + { \Delta p_i \cdot p_i  \over  m_N^2}  \Big( { \Delta p_i \cdot q \over q^2}   p_f \cdot q +{\Delta p_i \cdot p_f  }   \Big) \Big)  \nonumber  \\
& -  &  {2 \over q^2} \left( p_i^{\mu} q^{\nu}  + p_i^{\nu} q^{\mu} \right)  q \cdot p_f   - {2 \over q^2} (  p_f^{\mu} q^{\nu}   +p_f^\nu q^{\mu} )   \Big( q \cdot ( \Delta p_i + p_{i} ) + {\Delta p_i \cdot p_i \over  m_N^2}  \Delta p_i \cdot q \Big)  \nonumber  \\
&  - &  {2 \over q^2} \left(\Delta p_i^\mu q^{\nu}  + \Delta p_i^{\nu} q^{\mu} \right) q \cdot p_{f}  \Big( 1  + { \Delta p_i \cdot p_i \over  m_N^2}   \Big)  \Big]  
\nonumber \\
& + &   F_1 F_2 \kappa  \Big[  g^{\mu\nu}  \Big( {  \Delta p_i \cdot p_i \over  m_N^2} q \cdot ( 2 p_f - \Delta p_i ) - 2 \Delta p_i \cdot q \Big)  + q^{\mu}q^{\nu}  \Big(  { \Delta p_i \cdot p_i  \over m_N^2 q^2} q \cdot( \Delta p_i - 2 p_{f}) -   2    \Big) \nonumber \\
&  & \ \ \ \ \ \ \ \  - \left( p_i^{\mu} q^{\nu}  + p_i^{\nu} q^{\mu} \right)  + (  p_f^{\mu} q^{\nu}   +p_f^\nu q^{\mu} )  \Big] \nonumber \\
& + &  F_2^2  {\kappa^2 \over 2 m_N^2}    \Big[  g^{\mu\nu} \big[ \left( q^2 \ \Delta p_i \cdot p_f  - 2 \ q \cdot \Delta p_i \ q \cdot p_f\right) \Big( 1 + {\Delta p_i \cdot p_i \over  m_N^2 }   \Big) +q^2 \ \Delta p_i \cdot p_i  \big] \nonumber  \\
&  & \ \ \ \ \ \ \ -  \left( \Delta p_i^{\mu} p_f^{\nu}  + \Delta p_i^{\nu} p_f^{\mu} \right) q^2 \Big( 1 + {\Delta p_i \cdot p_i \over  m_N^2 }   \Big)  -  q^{\mu}q^{\nu}  \big[  \Delta p_i \cdot  p_f   \Big( 1 + {\Delta p_i \cdot p_i \over  m_N^2 }   \Big) -  \Delta p_i \cdot p_i  \big]  \nonumber  \\
&  &  \ \ \ \ \ \ \  +   \left( \Delta p_i^{\mu} q^{\nu}  + \Delta p_i^{\nu} q^{\mu} \right) q \cdot p_f  \Big( 1 + {\Delta p_i \cdot p_i \over  m_N^2 }   \Big) +   \left( p_f^{\mu} q^{\nu}  + p_f^{\nu} q^{\mu} \right) q \cdot \Delta p_i  \Big( 1 + {\Delta p_i \cdot p_i \over m_N^2 }   \Big)   \Big].
\end{eqnarray}

With our choice of reference frame (Fig.(\ref{sc_planes})), one can expand the $L_{\mu\nu}H^{\mu\nu}$ product  in the following form:
\begin{eqnarray}
L_{\mu \nu} H^{\mu \nu}_N & = & \left( L_{00}H^{00} - 2 L_{0z}H^{0z} + L_{zz}H^{zz} \right) + \left( -2 L_{0 \parallel }H^{0 \parallel } + 2 L_{z \parallel }H^{z \parallel} \right)  \nonumber \\
& + & \frac{1}{2} \left( L_{\parallel \parallel } + L_{\bot \bot} \right) \left( H^{\parallel \parallel } +  H^{\bot \bot} \right)  + {1 \over 2 } \left( L_{\parallel \parallel } - L_{\bot \bot} \right) \left( H^{\parallel \parallel } -  H^{\bot \bot} \right).
\end{eqnarray}
Furthermore,  using the gauge-invariance  of  leptonic current, one expresses the above product in the form:
\vspace{-0.3cm}
\begin{eqnarray}
L_{\mu \nu} H^{\mu \nu}_N & = & L_{00} \left( H^{00} - 2 {q^0 \over q_z } H^{0z} + \left({q^0 \over  q_z}\right)^2 H^{zz} \right) + 2 L_{0 \parallel } \left( -  H^{0 \parallel } + {q^0 \over q_z } H^{z \parallel} \right)  \nonumber \\
& + & \frac{1}{2} \left( L_{\parallel \parallel } + L_{\bot \bot} \right) \left( H^{\parallel \parallel } +  H^{\bot \bot} \right)  + {1 \over 2 } \left( L_{\parallel \parallel } - L_{\bot \bot} \right) \left( H^{\parallel \parallel } -  H^{\bot \bot} \right) \nonumber \\
& = & Q^2 (\tan(\theta/2))^2 \left( \eta_L V_{N,L} + \eta_{TL} V_{N,TL} \cos(\phi) + \eta_T V_{N,T} + \eta_{TT} V_{N,TT} \right).
\end{eqnarray}
Using the definitions of  $\eta_i$ for $ i=L,T,TL,TT$, from Eq.(\ref{eta}) for hadronic structure functions,  ($V_{N,i} $), one obtains:
\begin{eqnarray} \label{V-Gross-app}
	V^N_L & =& \frac{ {\bf q}^4 }{Q^4} \left( H^{00} - 2 {q^0 \over q_z } H^{0z} + ({q^0)^2 \over {\bf q}^2 } H^{zz} \right) = \frac{ {\bf q}^2 }{4 Q^2} \left( H^{++}{ Q^2 \over (q^+)^2 } + 2 H^{+-} + {(q^+)^2 \over  Q^2 } H^{- -} \right) \nonumber  \\
	V^N_{TL} & =&  2 { {\bf q}^2 \over Q^2 } \left( {q^0 \over q_z } H^{z \parallel}_N -  H^{0 \parallel }_N \right) = {|{\bf q}| \over q^+} \left( H^{+ \parallel}_N + H^{- \parallel }_N {(q^+)^2 \over Q^2} \right) \nonumber \\  
	V^N_T & =&  H^{\parallel \parallel }_N +  H^{\bot \bot} _N  \\
		V^N_{TT} & =& H_N^{\parallel \parallel} - H_N^{\bot \bot},   
\end{eqnarray}
where we have used, $ - q_z = |{\bf q}|$, as well as the relation between components of the nucleonic tensor in light-cone and Minkowski coordinates:
\begin{eqnarray}\label{H-components-LC}
H^{00} &=& { 1 \over 4 }(H^{++} + 2 H^{+-} + H^{- -})  \nonumber  \\
H^{0z} &=& {1 \over 4} (H^{++} - H^{- -}) \nonumber  \\
H^{zz} &=& {1 \over 4}  (H^{++} - 2 H^{+-} + H^{- -})  \nonumber \\   
H^{0 \parallel } &=& {1 \over 2} (H^{+ \parallel} + H^{- \parallel} ) \nonumber  \\
H^{z \parallel}  &=&  {1 \over 2} (H^{+ \parallel} - H^{- \parallel} ).
\end{eqnarray}

From Eqs.(\ref{HN-on}, \ref{V-Gross-app})
we compute  the  explicit forms of the structure functions. 
In the reference frame of Fig.(\ref{sc_planes}), they are given by:
\begin{eqnarray}
V_{L \ prop}^N  & = &    F_1^2 {\bf{q}}^2  \dfrac{\alpha_N \alpha_f}{\alpha_q^2} \Big(  {m^{2}_N+p_\textbf{T}^2  \over Q^2} \dfrac{\alpha_q^2 }
{\alpha_N \alpha_f} +  1    \Big) -   F_1 F_2 {\bf{q}}^2  {\kappa } \left(   {m^{2}_N+p_\textbf{T}^2  \over Q^2} 
\dfrac{\alpha_q^2 }{\alpha_N \alpha_f} +  1    \right)   \nonumber \\
&  & + F_2^2 {\bf{q}}^2  \left(  {\kappa \over 2 m_N}  \right) ^2    \left(  (m^{2}_N+p_\textbf{T}^2 )\dfrac{\alpha_q^2 }{\alpha_N \alpha_f} +  4 p_{\textbf{T}}^2  \right)  \nonumber \\ 
V_{L \ inst}^N  & =  &   F_1^2{   \alpha_{N} \over \alpha_{q}}  {\bf{q}}^2 \left(     1 - \left( {m^{2}_N+p_\textbf{T}^2  \over Q^2}  \dfrac{\alpha_q^2 }{\alpha_N \alpha_f}  \right)^2  
+  {(m^{2}_N+p_\textbf{T}^2 ) \over 2m^2} \dfrac{\alpha_q }{\alpha_N }   + \left( {m^{2}_N +p_\textbf{T}^2  \over Q^2} \dfrac{\alpha_q^2 }{\alpha_N \alpha_f} -  1    \right) ^2 \right)  \nonumber \\
	& & - 2 F_1 F_2 \kappa {\alpha_{N} \over \alpha_{q}}  {\bf{q}}^2 {   (q\cdot \Delta p_i)^2 \over m^2 Q^2} \left(    2 {\alpha_{f} \over \alpha_{q}}  + 2{  m^2 \over  q\cdot \Delta p_i   }  +1  \right) \nonumber   \\
	& & +   F_2^2 \Big( {\kappa \over m^2_N}  \Big)^2 {\bf{q}}^2 ~ q\cdot \Delta p_i  \left( 1 +   {q\cdot \Delta p_i  \over m^2 } { \alpha_N \alpha_f \over \alpha_q^2 }   \right) \nonumber \\
V_{TL \ prop}^N & = &    |{\bf{q}}|  \dfrac{ \alpha_N   +   \alpha_f }{\alpha_q} p_{\textbf{T}} \left( 2 F_1^2 + 2 F_2^2 \Big( {\kappa \over 2 m_N}  \Big)^2 Q^2\right)  \left( 1  +  {m^{2}_N+p_{\textbf{T}}^2 \over Q^2} \dfrac{ \alpha_q^2 }{\alpha_N \alpha_f} \right)   \nonumber \\
V_{TL \ inst}^N  & = &   8 |{\bf{q}}| { q \cdot \Delta p_i \over Q^2 }  p_\textbf{T}  \left( 1 +{p_i \cdot \Delta p_i  \over m^2} \right) \left(   F_1^2  +  F_2^2 \Big( {\kappa \over 2 m_N}  \Big)^2  \right)   \nonumber 	\\
V_{T \ prop \  \ }^N & =& \  F_1^2\left(   2 (m^{2}_N+p_{\textbf{T}}^2 )\dfrac{\alpha_q^2}{\alpha_N \alpha_f} +  4 (p_{\textbf{T}})^2     \right)  + 2 \kappa F_1 F_2   \left(  (m^{2}_N+p_{\textbf{T}}^2 ) \dfrac{\alpha_q^2}{\alpha_N \alpha_f} +  Q^2 \right)  \nonumber  \\ 
& & + F_2^2 \left(  {\kappa \over 2 m_N} \right) ^2 \left(  2 \dfrac{\alpha_N \alpha_f}{\alpha_q^2}  \left(  (m^{2}_N+p_{\textbf{T}}^2 ) \dfrac{\alpha_q^2}{\alpha_N \alpha_f} +  Q^2 \right) ^2  -  4 Q^2  p_{\textbf{T}}^2 \right)  \nonumber  \\
V_{T \ inst \ \ }^N &  =&   8 F_1^2   \left( q\cdot \Delta p_i  + p_f \cdot \Delta p_i  {p_i \cdot \Delta p_i \over m^2 } \right)   +  8  F_1 F_2 \kappa \left( 1 +{p_i \cdot \Delta p_i  \over m^2} \right) \left( q\cdot \Delta p_i  - p_f \cdot q  {p_i \cdot \Delta p_i \over m^2 + p_i \cdot \Delta p_i  } \right) \nonumber  \\
& & + 8 F_2^2 \left(  {\kappa \over 2 m_N}  \right) ^2  \left( 1 +{p_i \cdot \Delta p_i  \over m^2} \right) \left(  q \cdot p_f ~ q\cdot \Delta p_i  + Q^2 ~p_f \cdot \Delta p_i + Q^2  { m^2 ~p_i \cdot \Delta p_i \over m^2 + p_i \cdot \Delta p_i  } \right)     \nonumber  \\
V_{TT \ prop}^N  & = & 4 p_{\textbf{T}}^2 \left(  F_1^2 +   F_2^2  {\kappa^2 \over 4 m_N^2} Q^2   \right) \nonumber \\ 
V_{TT \ off}^N  & =  &  0.
\label{V_LF}
\end{eqnarray}

The kinematic variables, and scalar products used in the calculation are as follows:

The light-cone momentum fractions are:
\begin{equation}
\alpha_N= {2 p_N^+\over p_d^+}= {2 (E_N + p_{N, z}) \over p_d^+}, \ \ \  \alpha_q = {2 q^+\over p_d^+}= {2(q^0 - |{\bf{q}}|) \over p_d^+}, \ \ \ \alpha_f = \alpha_N + \alpha_q
\end{equation}
and the off-shell factor is, $ \Delta{p}_i^\mu=p^\mu_{i } - p^\mu_{i, on} $, with, $ p^\mu_{i }=p^\mu_{d}-p^\mu_{r} $. 
Since $\Delta p_i^+ = \Delta p_{i}^\perp = 0$, we have, $ 2 \Delta\sh{p}_i=\gamma^{+} \left( p^-_i - p^-_{i, on }\right) $ with the minus component 
defined as follows:
\begin{equation}
\Delta{p}_i^-=p_{i }^- - p_{i \ on}^- = -q^- + (p_f^- - p_{i \ on}^-) = \dfrac{Q^2}{q^+} -   \dfrac{m^{2}_N+p_\bot^2 }{p_f^+ p_i^+}  q^+ .
\end{equation}
The scalar products of initial ($p_{i,on}^\mu$), final ($p_{f}^\mu$) and transferred ($q^\mu$)   momenta   with the off-shell factor $\Delta{p}_i^\mu$, can be  written as:
\begin{eqnarray}
2 \Delta{p}_i \cdot p_i & = &   Q^2 \dfrac{\alpha_N}{\alpha_q}  -   (m^{2}_N+{\bf p_{T}}^2 ) \dfrac{\alpha_q}{\alpha_f }   \nonumber \\  2 \Delta{p}_i \cdot p_f& = &    Q^2 \dfrac{\alpha_f}{\alpha_q}  -   (m^{2}_N+{\bf p_{T}}^2 ) \dfrac{\alpha_q}{\alpha_i }\nonumber  \\  
2 \Delta{p}_i \cdot q & = &    Q^2  -   (m^{2}_N+{\bf p_{T}}^2 ) \dfrac{\alpha_q^2}{\alpha_f \alpha_N}.
\end{eqnarray}

\end{document}